\documentclass{article}

\usepackage{cite,latexsym,amsmath,amssymb,tabularx,supertabular,bm,wrapfig,subfigure}

\newcommand{\tr}{{\rm tr}}               

\usepackage{pstricks,color}
\definecolor{lightgray}{gray}{.80}
\newgray{verylightgray}{.80}

\newif\ifpdf
\ifx\pdfoutput\undefined
     \pdffalse                           
     
     \usepackage{graphicx}
\else
     \pdfoutput=1
     \pdftrue
      
     \usepackage[backref,colorlinks=true]{hyperref}
     \usepackage[pdftex]{graphicx}    
\fi

\begin{document}

\noindent{\Large\bf Non-global jet evolution at finite $N_c$}\\
  \vskip 0.5cm
  {\large Heribert Weigert
  }\\

  {\small
    Institut f\"ur theoretische Physik, 
          Universit\"at Regensburg, 93040 Regensburg,  Germany
   }
\vspace{.2cm}
\noindent\begin{center}
\begin{minipage}{.915\textwidth}
  {\small\sf Resummations of soft gluon emissions play an important
    role in many applications of QCD, among them jet observables and
    small $x$ saturation effects. Banfi, Marchesini, and Smye have
    derived an evolution equation for non-global jet observables that
    exhibits a remarkable analogy with the BK equation used in the
    small $x$ context. Here, this analogy is used to generalize the former
    beyond the leading $N_c$ approximation. The result shows striking
    analogy with the JIMWLK equation describing the small $x$
    evolution of the color glass condensate.  A Langevin description
    allows numerical implementation and provides clues for the
    formulation of closed forms for amplitudes at finite $N_c$. The
    proof of the new equation is based on these amplitudes with ordered soft
    emission. It is fully independent of the derivation of the JIMWLK
    equation and thus sheds new light also on this topic.  }
\end{minipage}
\end{center}
\vspace{1cm}

\section{Introduction}
\label{sec:introduction}

The starting point for the issues discussed and developed in this
paper is a beautiful exposition by Banfi, Marchesini, and Smye
(BMS)~\cite{Banfi:2002hw} of an evolution equation describing interjet
observables in hard QCD processes in which one asks for the energy
flow $E_{\text{out}}$ into a region away from all hard jets. With
$E_{\text{out}}$ much smaller than the hard scale of the jet, QCD
radiation into this region is small and these observables offer an
ideal opportunity to study nonperturbative effects. The central result
of~\cite{Banfi:2002hw} is an evolution equation that resums, for large
$Q/E_{\text{out}}$ and {\em in the large $N_c$ limit}, all leading
terms arising from large angle soft emission. In the following, I will
refer to this equation as the BMS equation. Subsequent analysis of the
consequences implied by this equation presented in the same
publication confirm features found using numerical methods by Dasgupta
and Salam~\cite{Dasgupta:2001sh, Dasgupta:2002bw}. The equation itself
is nonlinear in the part describing contributions arising {\em inside}
the jet regions and linear outside.  The contributions inside, quite
surprisingly, have a striking analogy with the Balitsky-Kovchegov (BK)
equation~\cite{Balitsky:1996ub, Balitsky:1997mk, Kovchegov:1999ua},
which resums small $x$ corrections in the nonlinear domain. There it
is used to describe saturation phenomena that occur in DIS at small
$x$, possibly off large hadronic targets. For small angle emission in
a certain frame, the analogy becomes even more striking: the
contributions of BMS inside the jet regions become identical to the BK
equation. That such an analogy, if it exists at all, is restricted to
the in-region, is not surprising: in the derivation of the BK equation
one assumes, in fact requires, the presence of non-linear effects {\em
  everywhere} and any geometric exclusiveness of the type imposed by
jets was simply never considered. What appeared to be more surprising
is the relation of jet physics to BFKL
dynamics~\cite{Kuraev:1977fs,Bal-Lip}, which dominates the BK equation
in the small density limit. This link has been established
in~\cite{Marchesini:2003nh} and indeed comes into play in the limit of
small angle emission in a judiciously chosen frame.

I will be concerned with a different question: the generalization of
the BMS equation to finite $N_c$ based on the fact that such is
already available on the BK side of the above analogy.

The BK equation is the limit of a more general equation, the JIMWLK
(Jalilian-Marian Iancu McLerran Weigert Leonidov Kovner)
equation~\cite{ Jalilian-Marian:1997xn, Jalilian-Marian:1997jx,
  Jalilian-Marian:1997gr, Jalilian-Marian:1997dw,
  Jalilian-Marian:1998cb, Kovner:2000pt, Weigert:2000gi,
  Iancu:2000hn,Ferreiro:2001qy}\footnote{The order of names was chosen
  by Al Mueller to give rise to the acronym JIMWLK, pronounced ``gym
  walk''.}, a functional equation equivalent to an infinite hierarchy
of coupled evolution equations known as the Balitsky
hierarchy~\cite{Balitsky:1996ub, Balitsky:1997mk}.

BK emerges from JIMWLK by means of a factorization assumption strongly
linked with the large $N_c$ limit. As both BK and BMS equations use
eikonalized soft emission techniques, both in a strongly ordered
setting, it is only natural that there will be a host of technical
similarities in both situations that may serve as a vehicle to derive
a consistent generalization of the BMS equation that goes beyond the
large $N_c$ limit. This derivation will be the goal of this paper.

In Sec.~\ref{sec:analogy} I will shortly present both the BMS and BK
equations and highlight their structural analogy. This will provide me
with a tentative identification of the transition probabilities used
in the BMS equation with averages of ``dipole type'' operators akin to
those appearing in the BK equation, albeit in a different kinematic
domain and geometrical setting.

Using this identification in Sec.~\ref{sec:from-jimwlk-bk}, I will
suggest a nonlinear evolution equation that generalizes the BMS
equation in the same way as JIMWLK generalizes the BK equation. From a
single equation for a single function one generalizes to a functional
equation, equivalent to an infinite hierarchy of equations for an
infinite tower of correlators.

I will identify real-emission and virtual correction parts of the
resulting functional Fokker-Planck equation and discuss the issues of
real-virtual cancellation and highlight the fact that this is in
direct correspondence with infrared finiteness.

Sec.~\ref{sec:an-equiv-lang} will provide the translation of the
resulting functional Fokker-Planck equation to an equivalent Langevin
description.This will allow to identify explicitly both real and
virtual contributions to the amplitudes in a closed form.
At the same time one gains a tool to study this evolution equation at
finite $N_c$ numerically, in analogy to what has been done already in
the JIMWLK case~\cite{Rummukainen:2003ns}.  It will also provide
insights into the structures that will be needed to derive the
underlying Fokker-Planck equation.

The remainder of the paper is devoted to a derivation of this equation
from first principles. Due to the functional nature of the equation,
the most economical way to proceed is to consider generating
functionals for the objects of interest. This will allow me to write
closed functional expressions for all the relevant amplitudes and
transition probabilities at {\em finite} $N_c$ in terms of a gluon
emission operator whose form is motivated by the comparison with
JIMWLK -- the amplitudes found are of course the relevant QCD
amplitudes of~\cite{Bassetto:1983ik,Fiorani:1988by}.

In Sec.~\ref{sec:ampl-strongly-order} I will define generating
functionals for amplitudes that describe the transition of any bare
jet configuration with an arbitrary but fixed number of hard particles
that emit any number of soft gluons $g_1 \ldots g_n$ in the strongly
ordered domain $\omega_1 \gg\ldots \gg\omega_n$, all softer than the
original set of hard particles. At this stage I will focus on real
emission. My initial example will be based on a bare $q\Bar q$
configuration.

Sec.~\ref{sec:trans-prob-from} is used to formulate generating
functionals for transition probabilities corresponding to the above
amplitudes that will then serve as the objects for which the evolution
equation is derived. 

This is done in Sec.~\ref{sec:evol-equat-soft} by considering
derivatives of the expressions for real emission with respect to the
phase space boundary in complete analogy to the method of BMS.
Virtual corrections are then included in the usual way with a
coefficient fixed to ensure real-virtual cancellation in the infrared.
The result is a proof of the equation suggested in
Sec.~\ref{sec:analogy} and a verification of the fact that it indeed
resums {\em all} finite $N_c$ corrections in the kinematic domain
under consideration.


Conclusions and a short discussion on possible cross fertilizations
between the two fields will conclude this paper in
Sec.~\ref{sec:conclusions}.

\section{The analogy of BMS and BK equations}
\label{sec:analogy}

To present the analogy, let me give a short account of both equations,
starting with BMS. I will closely follow the exposition
in~\cite{Banfi:2002hw}.

The observable considered by BMS is given by
\begin{equation}
  \label{eq:Sig-ee}
  \Sigma_{e^+e^-}(E_{\text{out}})=\sum_n\int\frac{d\sigma_n}{\sigma_T}\cdot
  \Theta\!\left(E_{\text{out}}-\sum_{h\,\in\,C_{\text{out}}}\omega_h\right)
  \ .
\end{equation}
Here $E_{\text{out}}$ is the sum of all energies of soft gluons
emitted into the region outside some predefined jet region, ${\cal
  C}_{\text{out}}$ and $d\sigma_n$ the $n$-hadron distribution and
$\sigma_T$ the total cross section. The dependence on jet geometry and
total energy $Q$ is understood. The process is dominated by iterative
soft emission from eikonalized gluons that are not deflected from their
original trajectory. The natural kinematic variables are thus the
directions of the emitters ($a,b,k,p,\bar p$ in what follows) and the
corresponding energies. Phase space integrals correspondingly will
involve integrals over solid angles and energies.

A Mellin transformation represents the cross section in terms of
transition probabilities $G_{ab}(E,E_{\text{out}})$
\begin{equation}
  \label{eq:Sig-ee1}
  \Sigma_{e^+e^-}(E_{\text{out}})\>=\>\int
\frac{d\nu\,e^{\nu E_\text{out}}}{2\pi i \nu}
  \>G_{p\bar p}(E,\nu^{-1})
\simeq G_{p\bar p}(E,E_{\text{out}})\>,
\end{equation}
were a saddle point approximation has been taken.

$G_{ab}(E,E_{\text{out}})$, or $G_{a b}(E)$ where the explicit
$E_{\text{out}}$ dependence is not needed, satisfies the BMS equation
\begin{equation}
  \label{eq:BMS-1}
\begin{split}
  E\partial_E\>G_{ab}(E)=\int \frac{d^2 \Omega_k}{4\pi}\,\Bar\alpha_s
  w_{ab}(k)\left[u(k)\,G_{ak}(E)\cdot G_{kb}(E)-G_{ab}(E)\right]
\ ,
\end{split}
\end{equation}
or equivalently
\begin{equation}
  \label{eq:BMS-2}
  \begin{split}
  E\partial_E\>G_{ab}(E)
  &=-E\partial_E\>R_{ab}^{(0)}(E)\cdot G_{ab}(E)\\
  &+\int \frac{d^2 \Omega_k}{4\pi}\,\Bar\alpha_s
  w_{ab}(k)\,u(k)\,\left[\,G_{ak}(E)\cdot G_{kb}(E)-G_{ab}(E)\,\right]
\ .
  \end{split}
\end{equation}
$\Bar\alpha_s:=\alpha_s N_c/\pi$, $R_{ab}^{(0)}(E)$ is called the
single log Sudakov radiator for bremsstrahlung emission
\begin{equation}
  \label{eq:Rad}
  R_{ab}^{(0)}(E)\>=\>\int_0^E\frac{d\omega}{\omega}
  \int\frac{d^2 \Omega_k}{4\pi}\,\Bar\alpha_s\,w_{ab}(k)\,
  [1-u(k)]\>=\>\Delta\cdot r_{ab}\>,
\end{equation}
and
\begin{equation}
  \label{eq:wab}
  w_{ab}(k)=\frac{(p_ap_b)}{(p_ak)(kp_b)}=
  \frac{1-\cos\theta_{ab}}{(1-\cos\theta_{ak})(1-\cos\theta_{kb})}
\end{equation}
is the soft emission kernel. Conventions here are adapted to lightlike
momenta and the energies, $\omega_p, \omega_q$ have been factored out
in $(p q)$ compared to the 4-vector product: $p.q=\omega_p \omega_q (p
q) = \omega_p \omega_q(1-\cos\theta_{p q})$.

The jet geometry is encoded in the definition of $u(k)$, which
emerges from a Mellin factorization of the energy constraint in
Eq.~\eqref{eq:Sig-ee}:
\begin{equation}
  \label{eq:Theta-fac}
  \Theta\Big(\!E_{\text{out}}\!-\!
  \sum_{i\in{\cal C}_{\text{out}}}\omega_i\!\Big)=
  \int\frac{d\nu\,e^{\nu E_{\text{out}}}}{2\pi i \nu}\, \prod_i u(k_i)\>,
  \quad   u(k)=\Theta_{\text{in}}(k)+e^{-\nu\omega}\Theta_{\text{out}}(k)\>,
\end{equation}
with the $\Theta$ functions having support inside and outside the jet
regions respectively. In fact, BMS show that to good accuracy one may
take $u(k)$ to restrict the phase space integrals to the outside
region for the Sudakov radiator term and to the inside region for the
remainder of Eq.~\eqref{eq:BMS-2}.

For more details on the ingredients as well as the physics of this
equation see~\cite{Banfi:2002hw, Dasgupta:2001sh, Dasgupta:2002bw,
  Bassetto:1983ik, Fiorani:1988by} and references therein.

At this point I want to highlight the strategy employed by BMS in the
derivation of Eq.~\eqref{eq:BMS-1}. This evolution equation was
derived from the knowledge of the structure of the real emission part
alone.  The contribution of the virtual corrections was included after
the fact using the requirement of real-virtual cancellation in the
infrared.

The real emission part of $G$ needed for this argument can be written
as
\begin{equation}
  \label{eq:GR-ee}
  G_{p\bar p}^{({\text{real}})}(E,E_{\text{out}})=1\!+\!
\sum_{n=1}^{\infty}\!\int\prod_{i=1}^n
\left\{\Bar\alpha_s \frac{d\omega_i}{\omega_i}\frac{d^2 \Omega_i}{4\pi}u(k_i)\,
\Theta(E\!-\!\omega_i)\!\right\} W_n(p k_{1} \ldots k_{n} \bar p)
\end{equation}
where the phase space of soft gluons is cut by $E$ and $W_n$ is the
large $N_c$ factorized version of the transition probability of a
color singlet into $(q\Bar q)_{\text{hard}} g^n_{\text{soft}}$.
\begin{equation}
  \label{eq:Wn}
  W_n(p k_{1} \ldots k_{n} \bar p)\>=\>\frac{(p\bar p)}
  {(pk_1)(k_1k_2)\ldots(k_n\bar p)}\>
\ .
\end{equation}
The line of argument then starts by taking a logarithmic derivative
$E\partial_E$ of Eq.~\eqref{eq:GR-ee} which uniquely fixes the
quadratic term in Eq.~\eqref{eq:BMS-1}. This term is affected by the
jet-geometry as signalled by the factor $u(k)$. The virtual
corrections are then taken into account by subtraction of a linear
term with a coefficient that ensures infrared finiteness through real
virtual cancellation. Virtual corrections occur globally, both inside
and outside the jet region, thus the factor $u(k)$ is absent. This
yields the second term in Eq.~\eqref{eq:BMS-1}.

That this indeed ensures infrared finiteness is best understood after
splitting off the Sudakov term as in Eq.~\eqref{eq:BMS-2}. Noting that
the integration in the Sudakov term is restricted to the out region
one recognizes that there is no danger of encountering any ill effects
from the poles in the kernel in this term. For the other term one
needs to prove (see~\cite{Banfi:2002hw}) that $G_{a a}(E) = 1$. Then
the quadratic and the linear terms cancel where the kernels diverge,
rendering the expression finite.

The BK equation on the other hand involves small $x$ kinematics in
which in the dipole approximation is taken. The latter corresponds to
the large $N_c$ limit of the above. While again one deals with hard
leading particles that emit softer ones in an eikonal manner, they all
share a common longitudinal direction, selected by the geometry of the
collision event. Instead of directions away from the hard creation
vertex of the jet(s), one is left with (2d) coordinates in the plane
transverse to the collision axis that characterize the location of the
hard emitters. To set the coordinate system in the small $x$ case, I
follow the common choice to represent the hard degrees of freedom as
given by eikonal lines along the $x^-$ direction (the direction of the
probe). The target, which is separated from the probe by a large
rapidity interval $\tau=\ln(1/x)$ (to leading accuracy it moves along
the $x^+$ direction) then supplies the field these eikonal lines
interact with. This plays a role similar to the soft emissions in the
BMS case. Any $x^+$ dependence is frozen out by time dilation. The
field variables then are given as 
\begin{equation}
  \label{eq:U-def}
  U_{\bm{x}} := P\exp\Big\{ -ig \int\limits_{-\infty}^{\infty}\! dz^- 
  A^+_{\text{soft}}(x^+=0,\bm{x},z^-)\Big\}
\end{equation}
where $A^+_{\text{soft}}(x^+=0,\bm{x},z^-)$ is the soft/Weiz\"acker
Williams field of the target. The basic object entering the DIS cross
section is the dipole cross section, the impact parameter integral of
\begin{equation}
  \label{eq:dipole-expect}
  N_{\tau; \bm{x y}} = \langle \Hat N_{\bm{x y}} \rangle_\tau 
\hspace{2cm}
 \Hat N_{\bm{x y}} = \tr(1-U^\dagger_{\bm{x}} U_{\bm{y}})/N_c
\ . 
\end{equation}
$N_{\tau; \bm{x y}}$ is the dipole function, $\Hat N_{\bm{x y}} $ the
dipole operator and the average $\langle \ldots \rangle_\tau$ is to be
thought to arise from an average of the $U$-configurations
characterizing the scattering with a given target at a given $\tau$.
The $\tau$ dependence is target independent and IR safe and hence
perturbatively calculable in form of an RG equation known as the BK
equation. As usual, the nonperturbative information is relegated to
its initial condition. To expose the similarity to the BMS equation it
is useful to write the BK equation in terms of
\begin{equation}
  \label{eq:S-def}
  S_{\tau; \bm{x y}} := \langle \Hat S_{\bm{x y}} \rangle_\tau 
\hspace{2cm}
 \Hat S_{\bm{x y}} = \tr(U^\dagger_{\bm{x}} U_{\bm{y}})/N_c 
\end{equation}
instead of  $N_{\tau; \bm{x y}}$. Then it takes
on the form
 \begin{equation}
    \label{eq:BK-S}
    \partial_\tau S_{\tau; \bm{x y}} 
    = 
    \frac{\alpha_s N_c}{2\pi^2} \int d^2z \ \Tilde{\cal K}_{\bm{x}
    \bm{z} \bm{y}}\,
  ( S_{\tau; \bm{x z}} S_{\tau; \bm{z y}} 
  -  S_{\tau; \bm{x y}} )
  \end{equation}
where 
\begin{equation}
  \label{eq:BK-K}
  \Tilde {\cal K}_{\bm{x}\bm{z} \bm{y}} := 
\frac{(\bm{x}-\bm{y})^2}{(\bm{x}-\bm{z})^2(\bm{z}-\bm{y})^2}
\end{equation}
is the BK-kernel.

Ignoring for the moment the Sudakov radiator term in
Eq.~\eqref{eq:BMS-2}, the similarity of Eqns.~\eqref{eq:BMS-2}
and~\eqref{eq:BK-S} is striking. A one to one relationship of
structures emerges if one maps directions onto transverse coordinates,
the kernels onto each other and tries to view $G_{a b}(E)$ as the
average of some $\tr(U^\dagger_a U_b)/N_c$ (where $a$ and $b$
represent the directions of the leading hard particles):
\begin{equation}
  \label{eq:interpretation}
  G_{a b}(E)
  \ \overset{{\text{\normalsize?}}}{\longleftrightarrow}\
  \langle \tr(U^\dagger_a U_b)/N_c \rangle_E 
  \ .
\end{equation}
Note that this is fully in line with the requirement that $G_{a
  a}(E)=1$ needed to ensure real virtual cancellation in the second
term of Eq.~\eqref{eq:BMS-2}.  While this appears to be quite
intuitive and fully in line with the physical interpretation it is not
clear how precisely to perform that latter part of the translation. One
would ask, for instance, how in detail to arrive at a consistent
definition of these $U$s and the averaging process suggested by
Eq.~\eqref{eq:interpretation} in the light of the tree like branching
process underlying the physics of the BMS equation. Additional
questions would be how to reconcile their role in probabilities with
that in amplitudes and many more in the same vein.

Despite these open questions, the analogy is too strong to ignore.
That the interpretation proposed in Eq.~\eqref{eq:interpretation}
should be possible suggests itself even more strongly, if one uses the
observation made by BMS that in the small angle emission limit in a
carefully chosen frame where the measure turns flat and the kernels
agree completely:
\begin{equation}
  \label{eq:w-small-angle}
  w_{a b}(k)  \to \Tilde {\cal K}_{{\Hat a \Hat k \Hat b}} 
\ .
\end{equation}
This is the region where Marchesini and
Mueller~\cite{Marchesini:2003nh} have established the link between jet
and BFKL-dynamics.

The benefit of demonstrating such a correspondence should be obvious
to any reader familiar with the color glass condensate (CGC): If there
is such an interpretation of the objects entering the BMS equation,
then there is hope one might generalize it along similar lines as the
BK equation, which can be viewed as the factorized limit of a more
general functional equation, the JIMWLK equation which I will turn to
next, in order to show how it reduces to BK under certain assumptions.
This will be the analogy used to suggest a generalization of the BMS
equation.

\section{From JIMWLK to BK: generalizing BMS by analogy}
\label{sec:from-jimwlk-bk}

To write the JIMWLK equation, consider the averaging procedure
entering the definition of $S_{\tau;\bm{x y}}$ to be performed using a
weight functional for eikonal lines $U$:
\begin{equation}
  \label{eq:average-def}
  \langle \ldots \rangle_\tau = \int \Hat D[U] \ldots \Hat Z_\tau[U]
\end{equation}
where $\Hat D[U]$ denotes a functional Haar measure, in keeping with
the group valued nature of the field variables $U$.

The JIMWLK equation describes $\tau$ dependence of the distribution
functional $\Hat Z_\tau[U]$. It was first written as a functional
Fokker-Planck with a very compact evolution operator
in~\cite{Weigert:2000gi} and reads
\begin{align}
  \label{eq:JIMWLK}
  \partial_\tau \Hat Z_\tau[U] = - H_{\text{FP}} \Hat Z_\tau[U]
\ . 
\end{align}
The Fokker-Planck Hamiltonian is defined as
\begin{subequations}
  \label{eq:JIMWLK-Hamiltonian}
\begin{align}
  \label{eq:JIMWLK-Hamiltonian-chi}
 H_{\text{FP}} := & \frac{1}{2}  \ i\nabla^a_{\bm{x}} 
 \chi^{a b}_{\bm{x y}}i\nabla^b_{\bm{y}}
\\
\label{eq:chidef}
\chi^{a b}_{\bm{x y}} := & 
 -\frac{\alpha_s}{\pi^2}\ \int\!\! d^2\!z\ {\cal K}_{\bm{x z y}} 
  \big[ (1-U^\dagger_x U_z) (1-U^\dagger_z U_y)\big]^{a b}
\end{align}
\end{subequations}
and can be shown to be positive definite.
In Eqns.~\eqref{eq:JIMWLK-Hamiltonian}, both a summation and an
integration convention is applied to repeated indices and coordinates.
Where
\begin{equation}
  \label{eq:JIMWLK-K}
  {\cal K}_{\bm{x z y}} = 
  \frac{(\bm{x}-\bm{z})\cdot(\bm{z}-\bm{y})}{(\bm{x}-\bm{z})^2
    (\bm{z}-\bm{y})^2}
\end{equation}
is the JIMWLK kernel and $i\nabla^a_{\bm{x}}$ is a functional
generalization of the left invariant vector field on the group
manifold [There is {\em no} coordinate derivative in the above]. Its
precise definition will be given shortly. Let me first provide two
structural comments:
\begin{itemize}
\item  Despite the apparent poles of the kernel at $\bm{z}=\bm{x},
\bm{y}$, the Hamiltonian will lead to explicitly finite evolution
equations because of the structure of the $U$-dependence 
which vanishes at the potentially singular point. This is also the
underlying reason for the cancellation of potential divergences in the
BK limit.
\item The equation is an equation for a distribution functional and
  thus a very compact way to represent an infinite number of (coupled)
  equations for individual correlators known as the Balitsky
  hierarchy~\cite{Balitsky:1996ub}. To recover members of this
  hierarchy, say the evolution equation for a correlator $O[U]$, one
  simply multiplies both sides of~\eqref{eq:JIMWLK} with $O[U]$ and
  the averages with~\eqref{eq:average-def}. The result then is an
  evolution equation for $\langle O[U] \rangle_\tau$
  \begin{equation}
    \label{eq:OU-evo}
    \partial_\tau \langle O[U] \rangle_\tau
    =
    - \langle H_{\text{FP}} O[U] \rangle_\tau
    \ .
  \end{equation}
  Because of the nonlinear nature of $H_{\text{FP}}$ the r.h.s. will
  not be expressible by $\langle O[U] \rangle_\tau$ alone but instead
  involves other correlators as well. The evolution equation of these
  new quantities will then also be needed, and the argument repeats
  itself, ultimetely leading to an infinite coupled hierarchy.
\end{itemize}
Returning to the left invariant vector fields I note that for present
purposes they may be taken to be defined as a variational derivative
according to
\begin{align}
  \label{eq:Lie-simpledef}
  i\nabla^a_{\bm{x}} := -[U_{\bm{x}} t^a]
  \frac{\delta}{\delta U_{\bm{x}, i j}}
  \ .
\intertext{$\frac{\delta}{\delta U_{\bm{x}, i j}}$ is the ordinary 
  functional (or variational) derivative w.r.t. the components of 
  the $U$ field:}
\frac{\delta}{\delta U_{\bm{x}, i j}} U_{\bm{y}, k l} 
 = \delta_{i k} \delta_{j l} \delta^{(2)}_{\bm{x y}}
\end{align}
where $ \delta^{(2)}_{\bm{x y}} := \delta^{(2)}(\bm{x}-\bm{y}) $ for
compactness. Operationally this then leads to
\begin{subequations}
  \label{Lie-der}
\begin{align}
  i\nabla^a_{\bm{x}} U_{\bm{y}} := & 
  -U_{\bm{x}} t^a \delta^{(2)}_{\bm{x y}}
  \ ,
  \hspace{1cm}
   i\nabla^a_{\bm{x}} U^\dagger_{\bm{y}} := 
    t^a U^\dagger_{\bm{x}} \delta^{(2)}_{\bm{x y}} \ .
\intertext{There is, of course a corresponding definition for the
right invariant vector fields $i\Bar\nabla^a_{\bm{x}}$:}
   i\Bar\nabla^a_{\bm{x}} U_{\bm{y}} := & 
   t^a U_{\bm{x}} \delta^{(2)}_{\bm{x y}}
   \ ,
  \hspace{1cm}
   i\Bar\nabla^a_{\bm{x}} U^\dagger_{\bm{y}} := 
   - U^\dagger_{\bm{x}} t^a \delta^{(2)}_{\bm{x y}}
   \ .
\end{align}
\end{subequations}
Their main properties are the commutation relations (which I display
leaving the functional nature aside for a second)
\begin{subequations}
\begin{align}
  \label{eq:comm-rules}
  [i\nabla^a,i\nabla^b] =  i f^{a b c} i\nabla^c 
  \hspace{1cm}
   [i\Bar \nabla^a,i\Bar \nabla^b] =  i f^{a b c} i\Bar \nabla^c  
   \hspace{1cm}
   [i\Bar \nabla^a,i \nabla^b] = & 0 
\ .
\end{align}
These then generate right and left translations
respectively:\footnote{Note that nevertheless, both correspond to
  what is called a left action of the group, as $f(e^{-i\omega}
  e^{-i\eta}, U) = f(e^{-i\omega} ,f(e^{-i\eta}, U))$ in both cases.}
\begin{align}
  \label{eq:translations}
  e^{-i\omega^a (i\nabla^a)} U = U e^{i\omega^a t^a} 
  \hspace{1cm}
  e^{-i\omega^a (i\Bar\nabla^a)} U =e^{-i\omega^a t^a} U 
\ .
\end{align}
Functional forms of this of course involve an integral in the
exponent:
\begin{align}
  \label{eq:translations-func}
  e^{-i\int_x\omega^a_x (i\nabla^a_x)} U_y = U_y e^{i\omega^a_y t^a}  
  \hspace{1cm}
  e^{-i\int_x\omega^a_x (i\Bar\nabla^a_x)} U_y =e^{-i\omega^a_y t^a} U_y 
\ .
\end{align}
$\nabla$ and $\Bar\nabla$  are interrelated by 
\begin{equation}
  \label{eq:nablarel}
  i\nabla^a_{\bm{x}} = 
  -[\Tilde U^\dagger_{\bm{x}}]^{a b} i\Bar\nabla^b_{\bm{x}};
  \hspace{1cm}
  i\Bar\nabla^a_{\bm{x}} =
  -[\Tilde U^\dagger]^{a b}_{\bm{x}} i\nabla^b_{\bm{x}}
\end{equation}
\end{subequations}
and ``representation conscious:'' With the above definitions for the
action on $U$ and $U^\dagger$ in the $q$ and $\Bar q$ representation,
it automatically follows from representation theory that acting
on $U$ or $U^\dagger$ in an arbitrary representation produces
analogous formulae with the generators appearing on the r.h.s. in that
representation.

Where there are different $U$ fields to distinguish, I will write
$\nabla^q_{U_{\bm{x}}}$ instead of $\nabla^a_{\bm{x}}$ for definiteness.

These definitions and properties will become the core technical tool
in the generalization of the BMS equation that is to follow below.
For the time being, let me note that using both of these objects,
$H_{\text{FP}}$ is elegantly written as
\begin{align}
  \label{eq:JIMWLK-Hamiltonian-2}
 H_{\text{FP}} = &
   -\frac{1}{2} \frac{\alpha_s}{\pi^2}\ {\cal K}_{\bm{x z y}}\ 
  \big[ i\nabla^a_x i\nabla^a_y+i\Bar\nabla^a_x i\Bar\nabla^a_y
  +\Tilde U_z^{a b}(i\Bar\nabla^a_xi\nabla^b_y
  +i\nabla^a_x i\Bar\nabla^b_y) \big]
\end{align}
[integration convention for $x, z, y$].  While the factorized form of
Eq.~\eqref{eq:JIMWLK-Hamiltonian} is most useful in a derivation of a
Langevin description of the evolution that allows a numerical
implementation, this second form is more economical in the derivation
of evolution equations for given correlators that follow as a
consequence from Eq.~\eqref{eq:JIMWLK}.

Among these, the equation for the two point operator $\Hat S_{\bm{x
    y}}$ is what I will turn to next in order to make contact with the
BK equation. Multiplying both sides of Eq.~\eqref{eq:JIMWLK} with
$\Hat S_{\bm{x y}}$ and taking the average over $U$ according to
Eq.~\eqref{eq:average-def}, one immediately arrives at
\begin{align}
  \label{eq:twopoint-JIMWLK}
  \partial_\tau \big\langle \Hat S_{\bm{x y}}\big\rangle_\tau = & 
  \frac{\alpha_s}{2\pi^2}
  \Big\langle\Big[{\cal K}^{(1)}_{\bm{u} \bm{z} \bm{v}}\Big(
  i\nabla^a_{{\bm{u}}}i\nabla^a_{{\bm{v}}}
 +
 i\Tilde\nabla^a_{{\bm{u}}}i\Tilde\nabla^a_{{\bm{v}}}\Big)
   + {\cal K}^{(2)}_{\bm{u} \bm{z} \bm{v}}\Big(\big[U_{\bm{z}}\big]^{a
   b}(i\Tilde\nabla^a_{{\bm{u}}}i\nabla^b_{{\bm{v}}}
 +i\Tilde\nabla^a_{{\bm{v}}}i\nabla^b_{{\bm{u}}})\Big)\Big]
\frac{\tr(U_{\bm{x}}U^\dagger_{\bm{y}})}{N_c} \Big\rangle_\tau 
\nonumber \\  = &
   \frac{\alpha_s}{2\pi^2}
  \Big\langle\Big[\big(
  2 {\cal K}^{(1)}_{\bm{x} \bm{z} \bm{y}} -
  {\cal K}^{(1)}_{\bm{x} \bm{z} \bm{x}} -{\cal K}^{(1)}_{\bm{y}
    \bm{z} \bm{y}}\big) 
\Big(
-2 
C_{\text{f}} \frac{\tr(U_{\bm{x}}U^\dagger_{\bm{y}})}{N_c}\Big)
\nonumber \\ & \hspace{.3cm}
+\big(
  2 {\cal K}^{(2)}_{\bm{x} \bm{z} \bm{y}} -
  {\cal K}^{(2)}_{\bm{x} \bm{z} \bm{x}} -{\cal K}^{(2)}_{\bm{y}
    \bm{z} \bm{y}}\big) 2
  \big[U_{\bm{z}}\big]^{a b}
  \frac{\tr(t^a U_{\bm{x}}t^b U^\dagger_{\bm{y}})}{N_c}
   \Big] \Big\rangle_\tau
\ .
\end{align}
In these expressions I have taken pains to label the contributions
from the terms with and without an additional factor of $\Tilde
U_{\bm{z}}$ by $^{(1)}$ and $^{(2)}$ for ease of reference. In the
JIMWLK case there is no distinction between the kernels in the two
cases and one finds that the linear combinations of JIMWLK kernels
just assemble into the BK kernel:
\begin{equation}
  \label{eq:BK-kernel-from-JIMWLK}
  \Tilde {\cal K}_{\bm{x} \bm{z} \bm{y}} = \big(
  2 {\cal K}^{(i)}_{\bm{x} \bm{z} \bm{y}} -
  {\cal K}^{(i)}_{\bm{x} \bm{z} \bm{x}} -{\cal K}^{(i)}_{\bm{y}
    \bm{z} \bm{y}}\big) 
\end{equation}
in both terms. 
To compare with the factorizing limit, one applies the Fierz identity
\begin{equation}
  \label{eq:Fierz}
  \big[U_{\bm{z}}\big]^{a b}
  \tr(t^a U_{\bm{x}}t^b U^\dagger_{\bm{y}}) =
  \frac{1}{2}\Big(\tr( U_{\bm{x}}U^\dagger_{\bm{z}})
           \tr( U_{\bm{z}}U^\dagger_{\bm{y}})
           -\frac{1}{N_c}\tr( U_{\bm{x}}U^\dagger_{\bm{y}})
  \Big)
\end{equation}
to rewrite Eq.~\eqref{eq:twopoint-JIMWLK} as
\begin{equation}
    \label{eq:pre-BK-S}
    \partial_\tau \langle \Hat S_{\bm{x y}}\rangle_\tau 
    = 
    \frac{\alpha_s N_c}{2\pi^2} \int d^2z \ \Tilde{\cal K}_{\bm{x}
    \bm{z} \bm{y}}\,\langle \Hat S_{\bm{x z}} \Hat S_{\bm{z y}} 
  -  \Hat S_{\bm{x y}} 
  \rangle_\tau 
  \ .
\end{equation}
This immediately reduces to Eq.~\eqref{eq:pre-BK-S} if one assumes
factorization in the nonlinearity  on the r.h.s. by replacing
\begin{equation}
  \label{eq:factorization}
  \langle \Hat S_{\bm{x z}} \Hat S_{\bm{z y}}  
  \rangle_\tau \to S_{\tau; \bm{x z}}\>  S_{\tau; \bm{z y}}  
  \ .
\end{equation}
Without this assumption, the evolution of the two point function
contains a three point function, which in turn will couple to yet
higher orders via an infinite hierarchy fully contained in
Eq.~\eqref{eq:JIMWLK}. The reason for this is the presence of $\Tilde
U_{\bm{z}}$ in Eq.~\eqref{eq:JIMWLK-Hamiltonian-2}. It is important to
note that the factorization assumption~\eqref{eq:factorization}
decouples and factorizes the whole hierarchy. Note that the real
emission term originates solely from the contributions labelled $(2)$
in Eq.~\eqref{eq:twopoint-JIMWLK} while the virtual corrections
receive contributions from both.

Since it would appear that the factorization assumption is closely
linked with the large $N_c$ limit, it would appear that if an
interpretation of $G_{a b}(E)$ in terms of averages over eikonal
factors in analogy with $S_{\tau; \bm{x y}}$ is possible, these
objects should satisfy an evolution equation that is in close analogy
to the JIMWLK equation. This is the premise on which I base the
conjecture for the finite $N_c$ generalization of the BMS equation. To
write it down I need
\begin{enumerate}
\item to use the correspondences of variables and kernels noted in the
  above. In this step it proves useful, if I slightly redefine the
  functional aspect of the invariant vector fields to produce $\delta$
  functions adapted to the structure imposed by the solid angle
  measure. To this end I will write
  \begin{align}
    \label{nabla-def}
    i\nabla^a_p U_p = t^a U_p \Bar\delta(p-q)
  \end{align}
  where $\Bar\delta(p-q)$ contains Jacobian factors such that
  \begin{align}
    \label{eq:bar-delta}
    \int\frac{d\Omega_k}{4\pi} \Bar\delta(p-q) f(q) = f(p)
    \ .
  \end{align}
  This will simplify expressions considerably.  
\item More importantly, I need to take into account the inside-outside
  distinction of the jet geometry by allowing the counterparts of
  ${\cal K}^{(i)}$ to be different for $i=1,2$. Matching can then be
  done on the BMS level by considering the evolution equation for the
  two point function suggested in~\eqref{eq:interpretation}. This is
  easily done and amounts to an algebraic exercise that fully
  determines both kernels from a rederivation of the BMS equation from
  its generalization via exactly the same steps displayed above for
  the JIMWLK/BK pair. Clearly both analogues of ${\cal K}^{(i)}$ will
  be proportional to $w_{p q}(k)$ with the proportionality factor
  $f^{(i)}(k)$ carrying the jet geometry via some $u(k)$
  dependence.\footnote{Note that the calculations are somewhat
    simplified by the fact that $w_{p p}(k) = 0$. The relation
    between kernels on the functional side to those on the BMS side as
    given by Eq.~\eqref{eq:BK-kernel-from-JIMWLK} for JIMWLK/BK pair,
    becomes a matter of simple proportionality.}
\end{enumerate}
The result of this exercise is a Fokker-Planck Hamiltonian I will dub
$H_{\text{ng}}$ (for ``non-global'').  I will present it in two forms,
to parallel the two versions of the BMS equation,~\eqref{eq:BMS-1}
and~\eqref{eq:BMS-2}. To this end I will introduce an additional
function $\Tilde f^{(2)}(k)$ that will serve as the coefficient of the
generalization of the Sudakov radiator term. One has
the following forms of the Hamiltonian:
\begin{subequations}  
\label{eq:H_ng-all}
\begin{align}
  \label{eq:H_ng-0}
  H_{\text{ng}} :=  &-\frac{\alpha_s}{2\pi}\ w_{u v}(k) \Big[
   f^{(1)}(k)\big( i\nabla^a_{{\bm{u}}}i\nabla^a_{{\bm{v}}}
 +i\Bar\nabla^a_{{\bm{u}}}i\Bar\nabla^a_{{\bm{v}}}\big)
 +f^{(2)}(k)\big[U_{\bm{k}}\big]^{a
   b}\big(i\Bar\nabla^a_{{\bm{u}}}i\nabla^b_{{\bm{v}}}
 +i\Bar\nabla^a_{{\bm{v}}}i\nabla^b_{{\bm{u}}}\big)\Big]
\intertext{in analogy with~\eqref{eq:BMS-1}, and}
  \label{eq:H_ng}
  H_{\text{ng}} :=  &-\frac{\alpha_s}{2\pi}\ w_{u v}(k) \Big[ 
  \Tilde f^{(1)}(k)\Big(
  i\nabla^a_{{\bm{u}}}i\nabla^a_{{\bm{v}}}
 +i\Bar\nabla^a_{{\bm{u}}}i\Bar\nabla^a_{{\bm{v}}}\Big)
   \nonumber \\ &+  
   f^{(2)}(k)\Big( i\nabla^a_{{\bm{u}}}i\nabla^a_{{\bm{v}}}
 +i\Bar\nabla^a_{{\bm{u}}}i\Bar\nabla^a_{{\bm{v}}}+\big[U_{\bm{k}}\big]^{a
   b}(i\Bar\nabla^a_{{\bm{u}}}i\nabla^b_{{\bm{v}}}
 +i\Bar\nabla^a_{{\bm{v}}}i\nabla^b_{{\bm{u}}})\Big)\Big]
\end{align}
\end{subequations}
to parallel~\eqref{eq:BMS-2}. In both cases $u,v$ and $k$ are
integrated over according to an ``integration convention'' that uses
$\frac{d\Omega_{p}}{4\pi}$ as its measure for any of the momenta.
Retracing the steps leading from the definition of the JIMWLK equation
to~\eqref{eq:pre-BK-S} by eye should make it obvious that indeed, the
separation of terms in Eq.~\eqref{eq:H_ng} is such that the first line
generates the Sudakov radiator and the second the nonlinear evolution
inside the jet cones.

A comparison of the result with BMS allows one to determine only the
leading $N_c$ part of the $f^{(i)}$. Already the $1/N_c$ corrections
are not controlled by the matching. To this accuracy the $f$ are $N_c$
independent:
\begin{subequations}
  \label{eq:tilde-cond}
\begin{align}
   f^{(2)}(k) = & u(k)\\
  f^{(1)}(k) = & 1 \\
  \Tilde f^{(1)}(k) = & (1-u(k))  
  \ .
\end{align}
\end{subequations}
Actually, it would be unnatural for the Fokker-Planck Hamiltonian to
contain any explicit $N_c$ dependence. If this were the case, it would
require very special circumstances for it to incorporate {\em all}
finite $N_c$ corrections to BMS in a consistent manner. For this
reason I take Eqns.~\eqref{eq:H_ng-all},~\eqref{eq:tilde-cond}
together with a Fokker-Planck equation of the form
\begin{equation}
  \label{eq:FP-ng}
  E\partial_E \Hat Z_E[U] = -H_{\text{ng}} \Hat Z_E[U]
\end{equation}
to define my conjecture for the finite $N_c$ generalization of the BMS
equation. This result will, of course, be derived independently below.

The most important feature of this equation is its infrared finiteness
which directly relates to the real virtual cancellations being
correctly encoded in Eq.~\eqref{eq:H_ng}. Considering the Sudakov term
in $H_{\text{ng}}$ one finds that just as in the BMS limit, the phase
space integration over $k$ is restricted to the outside region where
no hard particles are to be found. Therefore there is no divergence
from the poles in $w_{u v}(k)$. The second term may be recast in
analogy to Eq.~\eqref{eq:JIMWLK-Hamiltonian} and the same finiteness
argument used there clearly carries over to this situation. By looking
at the BMS limit one sees that this same cancellation of terms
corresponds to a real-virtual cancellation in the conventional sense.
On the functional side, real and virtual contributions arise from
different terms.  Contributions with an additional factor $\Tilde U_k$
correspond to real emission, the rest to virtual corrections.

The idea to interpret jet transition probabilities as correlators of
path-ordered exponentials in singlet projections properly takes into
account gauge invariance, just the same way as the formulation of the
JIMWLK equation does.  The equation suggested therefore would appear
to consistently encode the physics it is intended to cover with a set
of very strong constraints. From this perspective it would be rather
surprising would it not be possible to arrive at this result from
first principles.

\section{An equivalent Langevin description}
\label{sec:an-equiv-lang}

Translation of a Fokker-Planck equation to an equivalent Langevin
description is a textbook topic, that for the present case is only
complicated by the functional nature of the equation and the group
valued nature of variables involved. That such a step is possible in
the JIMWLK case has been first suggested in~\cite{Weigert:2000gi} and
worked out in detail in~\cite{Blaizot:2002xy}.

Analytical results from the Fokker-Planck formulation are generically
hard to come by, although some information on the fixed point
structure has been obtained in the case of the JIMWLK equation.  This
makes the alternative formulation even more useful, as it paves the way
for a numerical implementation of the evolution equation. This has
already been successfully carried out in the JIMWLK case
in~\cite{Rummukainen:2003ns}.

The Langevin formulation abandons the description in terms of weight
functionals $\Hat Z_\tau$ in favor of one in terms of ensembles of
fields which are governed by the corresponding Langevin equations.
Explicitly, to calculate any observable $O[U]$ of the fields $U$ one
writes
\begin{equation}
  \label{eq:ensemble-average}
  \langle O[U] \rangle_\tau = \int\!\Hat D[U] O[U] \Hat Z_\tau[U] 
  \approx 
  \frac{1}{N} \sum\limits_{U\in {\sf E}[\Hat Z_\tau]}  O[U]
\end{equation}
where, separately at each $\tau$, the sum is over an ensemble ${\sf
  E}[\Hat Z_\tau]$ of $N$ configurations $U$. Its members can be
thought of as created randomly according to the distribution $\Hat
Z_\tau$. Clearly, for $N\rightarrow\infty$, the ensemble and $\Hat
Z_\tau$ contain the same information.

The dynamics are then encoded in a Langevin equation, a finite
difference equation in evolution ``time'' $\tau$ that contains a
random force driven by a noise $\Xi$.

Prerequisites are (termwise) positive definiteness of the
Fokker-Planck Hamiltonian in question. Only then can one introduce
bounded noise integral -- if needed separately for each positive term.
If the Hamiltonian can be (termwise) separated into two conjugate
factors in analogy to the structure of
Eq.~\eqref{eq:JIMWLK-Hamiltonian}, any correlation effects can be
included in the Langevin equation. The noise can then be taken to be
(termwise) uncorrelated. To establish the structure one may expect for
the result, let me give the corresponding expressions for the JIMWLK
equation.  There, the Langevin equation schematically\footnote{This is
  a continuous time version of the equation that strictly speaking is
  not unique.  The path-integral derivation shown
  in~\cite{Blaizot:2002xy} and maybe more transparently in the
  appendix of~\cite{Rummukainen:2003ns} makes it clear that we are to
  take a ``retarded'' prescription here in which the derivative on the
  l.h.s\ is taken as a finite difference and the fields on the r.h.s.\ 
  are determined at the previous time step.} reads
\begin{equation}
  \label{eq:Langevin}
  \partial_\tau\, [U_{\tau;\bm{x}}]_{i j} 
  = [U_{\tau;\bm{x}} i t^a]_{i j} \Big[\int\!\! d^2y\, 
  {\cal E}_{\bm{x} \bm{y}}^{a b;k}[U_\tau] 
  \xi^{b,k}_{\tau;\bm{y}}+\Hat\sigma^a_{\bm{x}}[U_\tau]\Big]
\end{equation}
where
\begin{equation}
  {\cal E}^{ab;k}_{\bm{x}\bm{y}}[U_\tau] = \left(\frac{\alpha_s}{\pi^2}\right)^{1/2} 
  \frac{(\bm{x}-\bm{y})_k}{(\bm{x}-\bm{y})^2} 
    [ 1 - \Tilde U^\dagger_{\tau;\bm{x}} \Tilde U_{\tau;\bm{y}} ]^{ab} 
\end{equation}
is the ``square root'' of $\chi$, $\chi^{ab}_{\bm{x y}} = {\cal
  E}^{ac}_{\bm{x}\bm{z}} {\cal E}^{c b}_{\bm{z}\bm{y}}$. Factorization
is complete and no termwise split is necessary (or even possible
without violating positivity). $\sigma$ is given by
\begin{equation}
  \label{eq:newsigdef}
  \Hat\sigma^a_{\bm{x}}[U_\tau]:= \frac{1}{2}\nabla^b_{\bm{y}} 
\Hat\chi^{a b}_{\bm{x y}}= 
 i \big(\frac{1}{2}\frac{\alpha_s}{\pi^2}\int\!\!d^2z 
{\cal K}_{\bm{x z x}}\Tilde \tr( \Tilde t^a \Tilde U_{\tau;\bm{x}}^\dagger  
\Tilde U_{\tau;\bm{z}})\Big)
\ .
\end{equation}
The $\xi$ are independent Gaussian random variables with
correlators determined according to
\begin{equation}
  \label{eq:etacorr}
  \langle\ldots \rangle_\xi = \int\!D[\xi]\, (\ldots)\,
  e^{-\frac{1}{2}\xi \xi}
\ .
\end{equation}

Fortunately all of the prerequisites listed above are also met for
$H_{\text{ng}}$.  In order to achieve termwise factorization I first
separate off the $f^{(2)}$ terms and further treat the two
contributions proportional to $\Tilde f^{(1)}$ separately.  All of
these are positive due to the definition of $u(k)$:
\begin{equation}
  \label{eq:u-def}
  u(k)=\Theta_{\text{in}}(k)+e^{-\nu\omega} \Theta_{\text{out}}(k)
\end{equation}
ensures $0< u(k)<1$ and the same then for the coefficient functions in
Eq.~\eqref{eq:H_ng}.

Correspondingly one would write the Hamiltonian on the JIMWLK level as
\begin{align}
  \label{eq:H_ng-2}
  H_{\text{ng}} 
= &\frac{1}{2} i\nabla^a_{u}\chi^{a b}_{u v}i\nabla^b_{v}
\end{align}
where $u,v$ are integrated over with $\frac{d\Omega_u}{4\pi}\frac{
  d\Omega_v}{4\pi}$ and
\begin{align}
  \label{chi_ng}
  \chi^{a b}_{u v} = &
 \   - \int \frac{d\Omega_k}{4\pi} \frac{\alpha_s}{\pi} w_{u v}(k) \Big[
   \Tilde f^{(1)}(k) (1+U_u^\dagger U_v)
   + f^{(2)}(k)(1-U_u^\dagger U_k)(1- U_k^\dagger U_v)\Big]^{a b}
   \ .
\end{align}
Factorization of $\chi$ according to $\chi = {\cal E} {\cal
  E}^\dagger$ is achieved by defining a three component structure
representing the three separately positive terms alluded to before:
\begin{align}
  \label{eq:Edef}
  {\cal E}_{p k}^{a b; \mu} = & \sqrt{\frac{\alpha_s}{\pi}}
\frac{p^\mu}{p.k} \big\{
\sqrt{\Tilde f^{(1)}(k)} \delta^{a b},
\sqrt{\Tilde f^{(1)}(k)}[U^\dagger_p]^{a b},
\sqrt{  f^{(1)}(k) } (1-U_p^\dagger U_k)^{a b}
\big\}
\ .
\end{align}
Correspondingly one has to introduce independent white noise for all
of the components,
\begin{equation}
  \label{eq:noise}
  \Xi_k^{b;\mu} = \big\{(\xi^{(1)})_k^{b;\mu},
  (\xi^{(1')})_k^{b;\mu},
  (\xi^{(2)})_k^{b;\mu}\big\}
\ .
\end{equation}
The only thing left to cope with is the measure and the
$\delta$-functions in the noise correlator, which need to be such
that
\begin{equation}
  \label{eq:noisecond}
\big\langle \big(\int\frac{\Omega_k}{4\pi} 
{\cal E}_{p k}^{a c; \mu}  \Xi_k^{c;\mu} \big) 
\big(\int\frac{\Omega_l}{4\pi} 
{\cal E}_{q l}^{a d; \nu}  \Xi_l^{d;\nu} \big)\big\rangle 
= \chi_{p q}^{a b}
\ .
\end{equation}
Now the measure is $\frac{d\cos\theta d\phi}{4\pi}$ and one 
needs the correlators to read
\begin{equation}
  \label{eq:xicorr-1}
  \langle  (\xi^{i})_p^{a;\mu} (\xi^{j})_q^{b;\nu} \rangle 
  = 4\pi \delta(\cos\theta_p-\cos\theta_q) \delta(\phi_p-\phi_q) 
  \delta^{i j} 
  g^{\mu\nu}
\ .
\end{equation}
With these preparations and the identification $\tau \leftrightarrow
\ln E$ one has a Langevin equation that reads
\begin{equation}
  \label{eq:Jet-Langevin}
  \partial_\tau\, [U_{\tau;p}]_{i j} 
  = [U_{\tau;p} i t^a]_{i j} \Big[\int\! \frac{d\Omega_k}{4\pi}\, 
  {\cal E}_{p k}^{a b; \mu}[U_{\tau; p}]\ \Xi^{b; \mu}_{\tau; k}\Big]
\ .
\end{equation}
In comparison with the JIMWLK case, any sigma-terms vanish because
$w_{p q}(k)$ satisfies $w_{p p}(k)=0$.

Note again that any continuum notation is deceptive. The equation to
consider is really a finite difference equation and upon iteration
will allow for an interpretation of subsequent, ordered soft gluon
emission.  To this end it is instructive to rewrite the Langevin
equation in terms of functional derivatives:
\begin{equation}
  \label{eq:Jet-Langevin-func}
  \partial_\tau\, [U_{\tau; p}]_{i j} 
  = [U_{\tau; p} i t^a]_{i j}  \Big[\int\! \frac{d\Omega_k}{4\pi}\, 
  {\cal E}_{p k}^{a b; \mu}[U_{\tau; p}]\ \Xi^{b; \mu}_{\tau; k}\Big] 
  = 
   \Big\{\int\! \frac{d\Omega_q}{4\pi} \frac{d\Omega_k}{4\pi}\ 
  i {\cal E}_{p q}^{a b; \mu}[U_{\tau; p}]\ \Xi^{b; \mu}_{\tau; k}
  i\nabla^a_{U_{\tau; q}} \Big\} [U_{\tau; p}]_{i j}
\ .
\end{equation}
The operator in the last version contains two terms in its third
component, the component responsible for the in-region. Using a
somewhat more compact notation for the $\tau$ dependence they read:
\begin{equation}
  \label{eq:exp-real-virt}
   \int\! \frac{d\Omega_q}{4\pi} \int\! \frac{d\Omega_k}{4\pi}\, 
  \Big[i({\cal E}^3)_{p q}^{a b; \mu}\ \Xi^{b; \mu}_k
  i\nabla^a_q\Big]_{\tau'} 
  =  \int\! \frac{d\Omega_q}{4\pi}\frac{d\Omega_k}{4\pi}\,
  i\sqrt{\frac{\alpha_s}{\pi}} 
  \frac{p^\mu}{p.q} \Big[(i\nabla^a_p+ \Tilde U^{b a}_q i\Bar\nabla^b_p) \ 
  \Xi^{a; \mu}_k\Big]_{\tau'} 
\ .
\end{equation}
Taking into account the fact that the Langevin description is defined
only in a $\tau$-discretized sense, one might be tempted to conclude
that the terms containing factors $\Tilde U_q$ correspond to real
gluon emission, while the others would generate virtual corrections,
not only in this, but also the other components.

A formal solution to~\eqref{eq:Jet-Langevin} is
\begin{align}
   [U_{\tau, p}]_{i j} =  P_\tau   \exp \Big\{\int^\tau_{\tau_0} d\tau'
  \int\! \frac{d\Omega_q}{4\pi}
   \frac{d\Omega_k}{4\pi}\, 
  i \Big[{\cal E}_{p q}^{a b; \mu}\ \Xi^{b; \mu}_k 
  i\nabla^a_q\Big]_{\tau'} \Big\} [U_{\tau_0, p}]_{i j}
\ ,
\end{align}
which again is to be interpreted in a discrete sense.  To write this
expression I have again adapted the definition of the functional
derivatives to include a $\delta$-function in $\tau$:
$i\nabla^a_{\tau,p} U_{\tau' q} = -U_{\tau,p}t^a
\Tilde\delta(p-q)\delta(\tau-\tau')$.

This would have the interpretation of subsequent real emission with
virtual corrections correctly taken into account.

Instead of jumping to conclusions at this point, I will substantiate
these ideas with a derivation from first principles that follows the
strategy of BMS. For this one needs a formulation in terms of
functionals capable of handling finite $N_c$ corrections. This I will
turn to next. While doing so, a number of interesting structures, like
a closed, finite $N_c$ version of Eq.~\eqref{eq:GR-ee}
will emerge alongside similar expressions for the underlying
amplitudes.

\section{Amplitudes in the strongly ordered domain}
\label{sec:ampl-strongly-order}

In order to find a suitable starting point to go beyond the large
$N_c$ limit one has to go rather far back and start with a general
discussion of soft gluon amplitudes along the lines given already
in~\cite{Bassetto:1983ik,Fiorani:1988by}.

To help organize the argument, I will start with the definition of a
generating functional for the (tree level) amplitudes of (real) soft
gluon emission from a $q\Bar q$ pair in the strongly ordered region
\begin{equation}
  \label{eq:strong-ordering}
   \omega_{k_n} \ll \omega_{k_{n-1}} \ll \ldots \omega_{k_1} 
  \ll \omega_p < \omega_q 
\end{equation}
written as
\begin{align}
  \label{eq:simple-qqb-gen-func}
  {\sf A}_{p\,q}^{i j}[\Xi] :=\sum\limits_{n=0}^\infty & 
\int \frac{d\Omega_{k_1}}{4\pi}\ldots\frac{d\Omega_{k_n}}{4\pi} 
A(_{q p k_1 \ldots  k_n}^{ i j a_1 \ldots a_n}) 
\Xi_{k_1}^{a_1}\ldots \Xi_{k_n}^{a_n}
\ .
\end{align}
The amplitudes for $n$-gluon emission from a initial $q\Bar q$ pair
are denoted by $A(_{q p k_1 \ldots k_n}^{ i j a_1 \ldots a_n})$, where
the momenta $q, p, k_1 \ldots k_n$ and color indices $i, j, a_1 \ldots
a_n$ are explicitly listed, a corresponding set of Lorentz-indices
$\mu_1\ldots \mu_n$ is suppressed. They are isolated by $n$-fold
(functional) differentiation w.r.t. $\Xi$ at $\Xi=0$. At this stage,
$\Xi$ is just an external source, the relationship to the noise of the
Langevin description will become clear later.

The $A(_{q p k_1 \ldots k_n}^{ i j a_1 \ldots a_n})$ are known to
satisfy an iterative structure in which the ${\text{color singlet}}\to
q\Bar q g^{n+1}_{\text{soft}}$-amplitude follows by induction from the
corresponding amplitude with $n$ soft gluons. As a result, the
$0$-gluon term in Eq.~\eqref{eq:simple-qqb-gen-func} fully determines
the whole functional, although the explicit construction of the $n$-th
order expression with the tools available to date becomes more and
more cumbersome with growing $n$. The rules of the game have been
explicitly demonstrated in~\cite{Fiorani:1988by} and require no
recourse to the $1/N_c$ limit.

I will now try to establish a method that will allow me to give a
closed operator form of the full tower of amplitudes contained in
Eq.~\eqref{eq:simple-qqb-gen-func}. While explicit generation of a
term with arbitrary $n$ will still be cumbersome, the fact that I have
a closed mathematical expression instead of a prescription for an
iterative procedure will allow me to use it in calculations and
eventually in the derivation of the evolution equation.


For this purpose, a slight generalization of the
definition~\eqref{eq:simple-qqb-gen-func} is operationally somewhat
more useful in that it allows me to specify an explicit functional
form that casts the iteration step of~\cite{Fiorani:1988by} as a
simple functional operation. This operation will involve nothing more
complicated than the functional differentiation rules given in
Eq.~\eqref{Lie-der} and a suitably defined $n=0$ term to start the
process of generating the amplitudes.

The starting point is the definition 
\begin{align}
  \label{eq:U-qqb-gen-func}
  {\sf A}_{p\,q}^{i j}[U,\Xi] :=\sum\limits_{n=0}^\infty & 
\int \frac{d\Omega_{k_1}}{4\pi}\ldots\frac{d\Omega_{k_n}}{4\pi} 
A(_{q p k_1 \ldots  k_n}^{\Tilde i \Tilde j a_1 \ldots a_n}) 
\Tilde U_{k_1}^{a_1 b_1}\Xi_{k_1}^{b_1}\ldots 
\Tilde U_{k_n}^{a_n b_n}\Xi_{k_n}^{b_n}\ 
[U^\dagger_p]_{i \Tilde i} [U_q]_{\Tilde j j}
\ .
\end{align}
Here I have simply added a factor of $U$ in the appropriate
representation to each leg of the diagrams contained in the definition
of the ordered amplitudes. Diagrammatically the first few contributions read:
\begin{align}
  \label{eq:U-qqb-gen-func-graphically}
   {\sf A}_{p\,q}^{i j}[U,\Xi] = 
   \begin{minipage}[c]{1.5cm}
   \begin{center}
     \includegraphics[height=2cm]{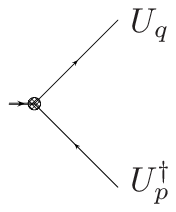}     
   \end{center}
 \end{minipage}
+
   \begin{minipage}[c]{2.5cm}
   \begin{center}
     \includegraphics[height=2cm]{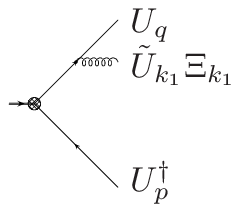}     
   \end{center}
 \end{minipage}
+
   \begin{minipage}[c]{2.5cm}
   \begin{center}
     \includegraphics[height=2cm]{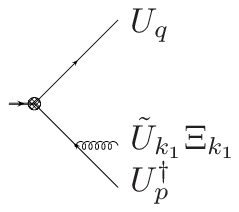}     
   \end{center}
 \end{minipage}
+   
\begin{minipage}[c]{2.5cm}
   \begin{center}
     \includegraphics[height=2cm]{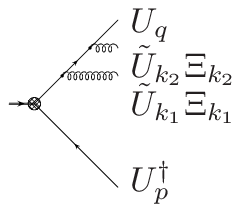}     
   \end{center}
 \end{minipage}
+
  \begin{minipage}[c]{2.5cm}
   \begin{center}
     \includegraphics[height=2cm]{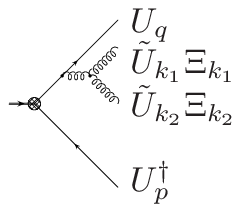}
   \end{center}
 \end{minipage}
+ \ldots
\end{align}
Obviously Eq.~\eqref{eq:simple-qqb-gen-func} is just the special case
at $U=1$: ${\sf A}_{p\,q}^{i j}[\Xi] = {\sf A}_{p\,q}^{i j}[1,\Xi]$.
The rationale behind this definition  lies in the very
nature of soft gluon emission underlying the construction of the
amplitudes. Per definition, soft gluon emission does not change the
direction of the parent (the emitter), in that role as a parent the
entity (be it a quark or a gluon) is therefore written as a path
ordered exponential along the direction of some momentum.  Emission of
a softer gluon, that in later steps will also serve as a parent will
then necessarily add an adjoint factor $\Tilde U_k$ and an eikonal
emission vertex $J_{l k}^\mu t^a$ (with $J^\mu_{p k} :=
\frac{p^\mu}{p.k}$), where the generator is to be taken in the
representation of the emitting object. As all these objects are
encoded as eikonal lines $U$, this will automatically occur if one
writes it via invariant vector fields $i{\Bar\nabla}^a_l$ just as in
the rewrite of the Langevin equation at the end of
Sec.~\ref{sec:an-equiv-lang}.  In functional form, one ends up with
the single emission operator
\begin{equation}
  \label{eq:real-emission-operator}
  \int\! \frac{d\Omega_l}{4\pi}
   \frac{d\Omega_k}{4\pi}\, 
  g\,J_{l k}^\mu \Tilde U^{a b}_k \ \Xi^{b; \mu}_k
  i{\Bar\nabla}^a_l
\ .
\end{equation}
Note how closely this resembles the real emission part of the Langevin
equation~\eqref{eq:Jet-Langevin}.  Eq.~\eqref{eq:U-qqb-gen-func}
should then be completely determined by the $n=0$ term in the sum
which describes $q\Bar q$-pair without additional soft gluons: higher
orders should just follow from repeated application
of~\eqref{eq:real-emission-operator}.  This needs now to be checked
against what is known about the soft gluon amplitudes.

First one needs to give an initial condition for this iteration, the
bare $q\Bar q$ term.  This reads
\begin{equation}
  \label{eq:qqb-iterative-initial}
 [{\sf A}^{(0)}]_{p\,q}^{i j}[U,\Xi] =
 \begin{minipage}[c]{2cm}
   \begin{center}
 \includegraphics[height=2cm]{bareqqb-U}     
   \end{center}
 \end{minipage}
:= 
M_2(q,p)  [U^\dagger_p U_q]_{i j} 
 \ ,
\end{equation}
and corresponds to a zero order amplitude of the form $A(_{q p}^{i j})
= M_2(q,p) \delta^{i j}$\footnote{For $M_n$ I have adopted the notation
of~\cite{Bassetto:1983ik,Fiorani:1988by}}.

The above prescription then claims that the $q\Bar q g_{\text{soft}}$
term is given by
\begin{equation}
  \label{eq:qqb-iterative-first}
 [{\sf A}^{(1)}]_{p\,q}^{i j}[U,\Xi]= \Big\{\int\! \frac{d\Omega_l}{4\pi} 
 \frac{d\Omega_k}{4\pi}\ g\,
  J_{l k}^\mu \Tilde U^{a b}_k \ \Xi^{b; \mu}_k
  i{\Bar\nabla}^a_l \Big\} [{\sf A}^{(0)}]_{p\,q}^{i j}[U,\Xi] 
\ .
\end{equation}
To demonstrate once how to use the machinery, I will go through the
steps explicitly. First insert $[{\sf A}^{(0)}]_{p\,q}^{i j}[U,\Xi] $
from~\eqref{eq:qqb-iterative-initial}. This yields
\begin{equation}
  \label{eq:qqb-iterative-first-1}
 [{\sf A}^{(1)}]_{p\,q}^{i j}[U,\Xi]= \Big\{\int\! \frac{d\Omega_l}{4\pi} 
 \frac{d\Omega_k}{4\pi}\ g\,
  J_{l k}^\mu \Tilde U^{a b}_k \ \Xi^{b; \mu}_k
  i{\Bar\nabla}^a_l \Big\} 
  M_2(q,p)  [U^\dagger_p U_q]_{i j} 
\ .
\end{equation}
All that is left to do is to use the differentiation
rules~\eqref{Lie-der} (with the adaptation to phase
space~\eqref{eq:new-nabla} taken into account) and differentiate the
two factors $U^\dagger_p$ and $U_q$. The variation eliminates the $l$
integral and one is left with
\begin{align}
  \label{eq:qqb-g-emission}
  [{\sf A}^{(1)}]_{p\,q}^{i j}[U,\Xi]= &
    \begin{minipage}[c]{2.5cm}
   \begin{center}
     \includegraphics[height=2cm]{qqb-gu-UXi}     
   \end{center}
 \end{minipage}
+
   \begin{minipage}[c]{2.5cm}
   \begin{center}
     \includegraphics[height=2cm]{qqb-gd-UXi}     
   \end{center}
 \end{minipage}
\nonumber \\ &
=\int\! \frac{d\Omega_k}{4\pi} 
  M_2(p,q) A_{p q}^\mu(k) [U^\dagger_p t^a U_q]_{i j} U^{a b}_k\Xi^{b; \mu}_k
=
   \begin{minipage}[c]{2.5cm}
   \begin{center}
     \includegraphics[height=2cm]{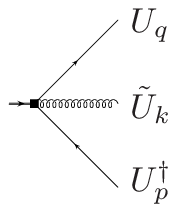}     
   \end{center}
 \end{minipage} \Xi_k
\end{align}
where
\begin{equation}
  \label{eq:A-def}
   A_{p q}^\mu(k) = J^\mu_{p k}-J^\mu_{q k}
   \ . 
\end{equation}
Now one isolates the amplitude via a variation in $\Xi$ and reads off
\begin{equation}
  \label{eq:qqb-g-amp}
  A(_{p q k}^{i j a}) =  M_2(p,q) A_{p q}^\mu(k) [ t^a ]_{i j} 
  =: M_3(p,k,q) [ t^a ]_{i j}
\end{equation}
with $ M_3(q,k,p)$ again as in the notation
of~\cite{Bassetto:1983ik,Fiorani:1988by}. Eq.~\eqref{eq:qqb-g-amp} in
fact is identical to Eq.~(40) of~\cite{Fiorani:1988by} and an
additional gluon emission, i.e. iteration with
Eq.~\eqref{eq:real-emission-operator}, leads to their Eq.~(42) for the
amplitude with $2$ soft gluons:
\begin{align}
  \label{eq:two-g-emission}
   [{\sf A}^{(2)}]_{p\,q}^{i j}[U,\Xi]= &
\begin{minipage}[c]{2.5cm}
   \begin{center}
     \includegraphics[height=2cm]{qqb-g2u-UXi}     
   \end{center}
 \end{minipage}
+
  \begin{minipage}[c]{2.5cm}
   \begin{center}
     \includegraphics[height=2cm]{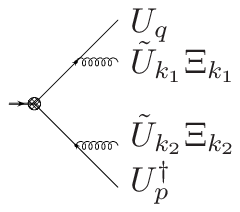}
   \end{center}
 \end{minipage}
+
  \begin{minipage}[c]{2.5cm}
   \begin{center}
     \includegraphics[height=2cm]{qqbggsplitu-UXi}
   \end{center}
 \end{minipage}
\nonumber \\ &
+
\begin{minipage}[c]{2.5cm}
   \begin{center}
     \includegraphics[height=2cm]{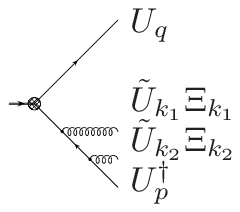}     
   \end{center}
 \end{minipage}
+
  \begin{minipage}[c]{2.5cm}
   \begin{center}
     \includegraphics[height=2cm]{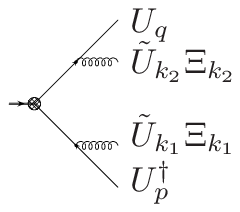}
   \end{center}
 \end{minipage}
+
  \begin{minipage}[c]{2.5cm}
   \begin{center}
     \includegraphics[height=2cm]{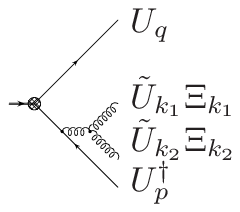}
   \end{center}
 \end{minipage}
\ .
\end{align}
(The color algebra for the explicit comparison is given in
App.~\ref{sec:color-algebra-induct}, case $n=1$.)  The pattern simply
follows the rules of functional differentiation and a comparison of
analytic expressions for the amplitudes as given by the iteration
rules of~\cite{Fiorani:1988by} was performed explicitly using symbolic
algebra tools up to $n=6$, where the task starts to become time
consuming. To compare beyond finite orders, a few additional tools are
required.

In the above I have implemented the ordering ``by hand,'' simply by
requesting the hierarchy~\eqref{eq:strong-ordering} for the
$\omega_{k_i}$. To write the all orders expression advertised
above, I need one more ingredient: the addition of an energy integral
to the emission operator~\eqref{eq:real-emission-operator} and an
appropriate extension of the definition of $i\Bar\nabla^a_p$ to
$i\Bar\nabla^a_{\omega_p,p}$ defined such that
\begin{equation}
  \label{eq:new-nabla}
  i\Bar\nabla^a_{\omega_p,p} U_q =  t^a U_p \Tilde\delta(p-q)
\end{equation}
where $\Tilde\delta(p-q)$ is adapted to the measure used in the
momentum and energy integrations. The natural choice for the latter is
to replace the solid angle integrations
in~\eqref{eq:real-emission-operator} by full fledged phase space
integrations:
\begin{equation}
  \label{eq:now-phase-space}
  \frac{d\Omega_k}{4\pi} \to d\Phi_k:=\frac{d^4k}{(2\pi)^4} 
  \delta(k^2)\theta(\omega_k) 
  = \frac{\omega_k d\omega_k}{(2\pi)^2} 
  \frac{d\Omega_k}{4\pi}\theta(\omega_k)
\ .
\end{equation}
$\Tilde\delta(p-q)$ is then defined as the appropriate $\delta$
function on the forward light cone, i.e. with Jacobian factors such that
\begin{equation}
  \label{eq:delta}
  \int  d\phi_k \Tilde\delta(p-k)f(k) = f(p) 
\ .
\end{equation}
(Note again the similarity of the roles played by $\tau$-ordering in
the Langevin description and energy ordering in this case.)  With
these definitions it is straightforward to verify that a replacement
of the solid angle integrations in the above by phase space
integrations leads to no change in the previous results, if one
implements energy ordering for subsequent definitions with ordering
$\theta$-functions, just as one would for the contour parameter in
case of path ordered exponential. To leading logarithmic accuracy this
will not change the result for the evolution equation.\footnote{Also
  the results below for the structure of the multiple ordered soft
  emission can be obtained both ways -- to have a formulation with all
  ingredients (including the ordering) written explicitly is merely a
  convenience from this perspective. What the formalism does, is to
  allow to give a closed expression for the generating functional,
  Eq.~\eqref{eq:exp-amp}.}

By now it should appear to be the natural choice if I write for the
generating functional (again suppressing the $\omega$-label for compactness)
\begin{align}
  \label{eq:exp-amp}
  {\sf A}_{p\,q}^{i j}[U,\Xi] 
= &
  P_{\omega_k}  \exp \Big\{\int\! d\Phi_l
   d\Phi_k\ 
  g\, J_{l k}^\mu \Tilde U^{a b}_k \ \Xi^{b; \mu}_k 
  i{\Bar\nabla}^a_l \Big\} 
[{\sf A}^{(0)}]_{p\,q}^{i j}[U,\Xi]  
%
\end{align}
where $P_{\omega_k}$ implements the strong ordering of
Eq.~\eqref{eq:strong-ordering}. I should actually use a notation that
displays the dependence on the functional form supplied in the zero
emission part and for instance write ${\sf A}_{p\,q}^{i
  j}[U,\Xi,M_2(p,q)]$.

In analogy with path ordered exponentials, this implies that the
exponential series is to be interpreted as
\begin{align}
  \label{eq:ordered-exp-series}
  P_{\omega_k} &  \exp \Big\{\int\! d\Phi_k d\Phi_l\ 
  g\,J_{l k}^\mu \Tilde U^{a b}_k \ \Xi^{b; \mu}_k
  i{\Bar\nabla}^a_l \Big\} 
\nonumber \\ := & 
\sum\limits_{n=0}^\infty   P_{\omega_k} \Big[ 
\Big\{\int\!d\Phi_{k_n} d\Phi_{l_n} \ 
  g\, J_{l_n k_n}^{\mu_n} \Tilde U^{a_n b_n}_{k_n} \ 
  \Xi^{b_n; \mu_n}_{k_n}
  i{\Bar\nabla}^{a_n}_{l_n} \Big\} 
\ldots
\Big\{\int\! d\Phi_{k_1} d\Phi_{l_1}\ 
 g\, J_{l_1 k_1}^{\mu_1} 
 \Tilde U^{a_1 b_1}_{k_1} \ \Xi^{b_1; \mu_1}_{k_1}
  i{\Bar\nabla}^{a_1}_{l_1} \Big\} 
\Big]
\end{align}
where the $1/n!$ are absent due to the explicit ordering of the
subsequent soft emissions.
To allow for more compact formulae below, I will introduce the notation
\begin{equation}
  \label{eq:emission-tree}
  {\sf U}[\Xi,U] :=  P_{\omega_k}   
  \exp \Big\{\int\! 
  d\Phi_k d\Phi_l\ 
  g\, J_{l k}^\mu \Tilde U^{a b}_k \ \Xi^{b; \mu}_k
  i{\Bar\nabla}^a_l \Big\} 
\end{equation}
for this functional differential operator.  The explicit mention of
the $U$-field in the notation will be needed later when different
types of eikonal fields will appear.  ${\sf U}[\Xi,U]$ will
turn out to have a meaning independent of the bare ($n=0$) term it is
used to act on. It will be called the (real) shower operator as it
will show up wherever one is forming that ordered soft gluon cloud
around a hard seed.
  
With these tools a direct comparison with the iterative scheme
of~\cite{Fiorani:1988by} at all orders becomes possible. There it was
shown (c.f. Eq.~(45) of~\cite{Fiorani:1988by}) that the $n$-th order
amplitude is of the form
\begin{equation}
  \label{eq:fioriani-n-step}
A(_{q p k_1 \ldots  k_n }^{i j a_1 \ldots a_n} ) 
= 
\sum\limits_{\Pi_{n+2}(l)} M_n(q,k_{l_1},\ldots, k_{l_n} ,p)
[t^{l_1}\ldots
 \ldots t^{l_n}]_{i j}
\end{equation}
where $\Pi_{n+2}(l)$ denotes the set of permutations of the $\{p,l_1,
\ldots, l_n,q\}$. Moreover, the amplitudes satisfy the recursion
relation
\begin{equation}
  \label{eq:iteration-goal}
A(_{q p k_1 \ldots  k_n k_{n+1}}^{i j a_1 \ldots a_n a_{n+1}}) 
=  \sum\limits_{j=1}^n \sum\limits_{\Pi_{n+2}(l)} 
M_n(q,k_{l_1},\ldots, k_{l_n} ,p)
A_{l_j l_{j+1}}(k_{n+1}) 
[t^{l_1}\ldots t^{l_j} t^{n+1}  
t^{l_{j+1}}  \ldots t^{l_n}]_{i j}
\end{equation}
(c.f. Eq.~(48) of~\cite{Fiorani:1988by}). This determines all orders,
once the first term is specified.

Since agreement has already been established for the first few terms
($n=0,1,2$ were compared explicitly, terms including $n=6$ have been
checked), all that is left to show, is that the functional
prescription given above leads to the recursion
relation~\eqref{eq:iteration-goal}. This is readily achieved: Assuming
the $n$-th order term, $[{\sf A}^{(n)}]_{p q}^{i j}[U,\Xi]$ to have
the form~\eqref{eq:fioriani-n-step}, one simply uses a single soft
gluon emission operator to find $[{\sf A}^{(n+1)}]_{p q}^{i
  j}[U,\Xi]$:
\begin{align}
  \label{eq:qqb-func-induction-step}
  [{\sf A}^{(n+1)}]_{p\,q}^{i j}[U,\Xi]  
= & \Big\{\int\!d\Phi_{k_{n+1}} d\Phi_{l_{n+1}}\
  g\,
  J_{l_{n+1} k_{n+1}}^{\mu_{n+1}} \Tilde U^{a_{n+1} b_{n+1}}_{k_{n+1}} \ 
  \Xi^{b_{n+1}; \mu_{n+1}}_{k_{n+1}}
  i{\Bar\nabla}^{a_{n+1}}_{l_{n+1}} \Big\} 
 [{\sf A}^{(n)}]_{p\,q}^{i j}[U,\Xi]
\nonumber \\ = &
\Big\{\int\! 
d\Phi_{k_{n+1}} d\Phi_{l_{n+1}}\
  g\,J_{l_{n+1} k_{n+1}}^{\mu_{n+1}} \Tilde U^{a_{n+1} b_{n+1}}_{k_{n+1}} \ 
  \Xi^{b_{n+1}; \mu_{n+1}}_{k_{n+1}}
  i{\Bar\nabla}^{a_{n+1}}_{l_{n+1}} \Big\} 
\nonumber \\ & \hspace{1cm}
\int 
d\Phi_{k_1} \ldots d\Phi_{k_n}
A(_{q p k_1 \ldots  k_n}^{\Tilde i \Tilde j a_1 \ldots a_n}) 
\Tilde U_{k_1}^{a_1 b_1}\Xi_{k_1}^{b_1}\ldots 
\Tilde U_{k_n}^{a_n b_n}\Xi_{k_n}^{b_n}\ 
[U^\dagger_p]_{i \Tilde i} [U_q]_{\Tilde j j}
\ .
\end{align}
It is straightforward to carry out the differentiation: The product
rule leads to a sum over $j$ similar to that in
Eq.~\eqref{eq:iteration-goal}. Then one uses $[\Tilde
t^{a_{n+1}}]^{b_{n+1} a_n} [t^{a_n}]_{k l} = [t^{n+1},t^{a_n}]_{i j}$
to combine the $J$ factors into the $A_{l_j l_{j+1}}(k_{n+1})$. After
this rearrangement the sum only runs over $j=1,\ldots,n$ instead of
$p,1,\ldots,n,q$ and matches~\eqref{eq:iteration-goal}. The algebraic
details of this rearrangement are given in
App.~\ref{sec:color-algebra-induct}.  This shows that one indeed
creates the same amplitudes with the {\em full} color structure, i.e.
without any recourse to the $1/N_c$ approximation.

While in the above, I have done nothing more but to provide a
functional form of a (real) ordered soft emission amplitude, I will
now show that there is structure contained in the above that is very
similar to the information contained in the Langevin equation and will
eventually lead to the RG equation stated earlier. The main
observation in this regard is that, as a consequence of the strong
ordering, any $\frac{\delta}{\delta \Xi^{a; \mu}_{k}}$ in which $k$ is
at (or near) the phase space boundary, there will be only a
contribution from the first emitted gluon ($n=1$). As a consequence,
for such variations one gets
\begin{align}
    \label{hard-variation}
    \frac{\delta}{\delta \Xi^{b; \mu}_{k}} 
    {\sf U}[\Xi,U]
={\sf U}[\Xi,U]
\Big\{\int\! d\Phi_l\
  J_{l k}^{\mu} \Tilde U^{a b}_{k} 
  i{\Bar\nabla}^a_l \Big\}
\ .
  \end{align}
Note that 
\begin{align}
    \label{eq:qqbg}
     {\sf A}_{p\,q\, k}^{i j b}[U,\Xi,M_3(p,k,q)] := 
 {\sf U}[\Xi,U]
\ 
  \Tilde U^{a b}_{k} [ U^\dagger_p t^a U_q]_{i j}
  M_3(p,k,q)
\end{align}
is the $q\Bar q g$ counterpart to Eq.~\eqref{eq:exp-amp}. In short,
Eq.~\eqref{hard-variation} states, that under variation at the phase
space boundary, one relates the generating functional for a given
tower of amplitudes to that of a tower with an additional hard gluon.
For the above example it relates the amplitudes with a hard $q\Bar q$
to those with hard $q\Bar q g$ content:
\begin{align}
    \label{eq:qqb-qqbg-rel}
     \frac{\delta}{\delta \Xi^{b; \mu}_{k}}
     {\sf A}_{p\,q}^{i j}[U,\Xi,M_2(p,q)] 
     =    {\sf A}_{p\,q\, k}^{i j b}[U,\Xi, A_{p\, q}^\mu(k)M_2(p,q)]
\ ,
\end{align}
or diagrammatically
\begin{equation}
  \label{eq:Xivar}
      \frac{\delta}{\delta \Xi^{b; \mu}_{k}}\> 
    {\sf U}[\Xi,U] \>
   \begin{minipage}[c]{2cm}
   \begin{center}
     \includegraphics[height=2cm]{bareqqb-U}     
   \end{center}
 \end{minipage} 
=\  
{\sf U}[\Xi,U]    \begin{minipage}[c]{2.5cm}
   \begin{center}
     \includegraphics[height=2cm]{qqbg-U}     
   \end{center}
 \end{minipage} 
\ .
\end{equation}
These two items, the ordered nature leading to~\eqref{hard-variation}
under ``hard'' variations, and the fact that in this situation the
number of hard legs of the soft emission amplitudes is increased by
one as exemplified in~\eqref{eq:Xivar}, will be the core observations
behind the evolution equation I am aiming at.

\section{Transition probabilities from $\Xi$ averages}
\label{sec:trans-prob-from}

As a first step to understand how to translate the above into
expressions for transition probabilities, let me take the situation of
a (real) $q\Bar q g^n_{\text{soft}}$ soft emission amplitude in which one
obtains the standard result
\begin{align}
  \label{eq:qqbgn-prob}
A(_{q p k_1 \ldots  k_n }^{i j a_1 \ldots a_n} ) 
[ A(_{q p k_1 \ldots  k_n }^{i j a_1 \ldots a_n} ) ]^\dagger
= &
\sum\limits_{\Pi_{n+2}(l)}\sum\limits_{\Pi_{n+2}(l')}
M_n(q,q_{l_1},\ldots, q_{l_n} ,p)
M_n(q,q_{l_1},\ldots, q_{l_n} ,p)^*\nonumber \\ & \hspace{2cm}
\times \tr(t^{a_{l_1}}\ldots\ldots t^{a_{l_n}} t^{a_{l'_n}}\ldots
 \ldots t^{a_{l'_1}})
\ ,
\end{align}
in which I have labelled the gluon momenta in the final state by
$q_1, \ldots, q_n$.  The same expression is obtained, if one considers
the $\Xi$ averaged product of functionals
\begin{align}
  \label{eq:prob-from-func-fixed}
  \langle  [{\sf A}^{(n)}]_{p\,q}^{i j}[U,\Xi] 
  [{\sf A}^{(n)}]_{p\,q}^{i j}[U,\Xi]^\dagger \rangle_{n,\Xi}
\end{align}
if the average is defined with a Gaussian distribution for $\Xi$ in
which only strongly ordered modes $q_1, \ldots, q_n$ have support. I
write
\begin{equation}
  \label{eq:xi-average}
  \langle \ldots \rangle_{n\Xi} := 
  \det(M_n)^{\frac{1}{2}} \int\!D[\Xi] \ldots 
  e^{-\frac{1}{2}\int d\phi_p d\phi_q \Xi_p^t M^{-1}_{n,\, p q} \Xi_q}
\end{equation}
where $d\phi_k$ denotes the phase space integral for momentum $k$ and
I have suppressed discrete indices. $M$ defines the $\Xi$ correlator,
which in the fully exclusive case considered above would read
\begin{equation}
  \label{eq:xicorr-2}
  M_{n,\, p q}^{a,\mu\ b,\nu} = 
\langle \Xi_p^{a, \mu} \Xi_q^{b, \nu}\rangle_{n,\Xi} 
= \delta^{a b} g^{\mu\nu} \Tilde \delta(p-q)\ 
  \sum\limits_{i=1}^n \Tilde\delta(q-q_i)
\end{equation}
In this case, because of the strong ordering of momenta in the
amplitudes, the $\Xi$ integral will strictly {\em pair} off the one
available momentum $k_i$ in the amplitudes falling into the same range
as $q_i$.

As such, this is only true if I have ordering in the emission
vertices, either done by hand in the 2-d version, or via
$P_{\omega_k}$ in the 2+1-d formulation. Only then, unwanted cross
terms are excluded.  The $U$ factors just cancel trivially.

To make contact with the non-global observables of the BMS setting,
one needs to depart from the fixed $n$ situation and replace the
$\sum\limits_{i=1}^n \Tilde\delta(q-q_i)$, which selects a given set of
final state momenta in the case where the number of final state gluons
is fixed, by a cutoff criterion that restricts real emission by
geometry and energy. By writing
\begin{align}
  \label{eq:M-BMS}
   M_{p q}^{a,\mu\ b,\nu} = 
\langle \Xi_p^{a, \mu} \Xi_q^{b, \nu}\rangle_\Xi  
=
\delta^{a b} g^{\mu\nu} \Tilde \delta(p-q)\ \theta(E-\omega_q) u(q)
\end{align}
one can indeed go beyond a final state with a fixed number of gluons
and consider
\begin{align}
  \label{eq:prob-from-func-all}
  \langle  {\sf A}_{p\,q}^{i j}[U,\Xi] 
  {\sf A}_{p\,q}^{i j}[U,\Xi]^\dagger \rangle_\Xi
\ ,
\end{align}
which will provide the finite $N_c$-generalization of $G_{a
  b}^{(\text{real})}$ of BMS.

In fact any degree of inclusiveness  may be imposed by modifying the
restrictions on the allowed modes appearing on the r.h.s. of
Eq.~\eqref{eq:M-BMS}.

At this point a few comments about the meaning of $\Xi$ as it appears
in the above are in order. Viewed from a diagrammatic point of view,
the $\Xi$ stand for nothing else but the final state soft gluons. This
explains why there is an additional factor of $i$ together with the
use of the free gluon phase space measure in the averaging, compared
to structures encountered in Sec.~\ref{sec:an-equiv-lang}. There,
$\Xi$ should be viewed simply as a perturbative gluon and the same
ingredients as those encountered here would show up in the formulation
of $S$-matrix elements via the amputation of external legs.

This realization should make it clear how to include virtual
corrections right from the outset, by a fairly simple modification of
the above. I will refrain from doing so here and instead keep a closer
parallel with the strategy employed
by~\cite{Banfi:2002hw,Bassetto:1983ik,Fiorani:1988by} and use the
expressions for real emission to deduce the structure of the evolution
equation.

\section{Evolution equations for soft semi-inclusive quantities}
\label{sec:evol-equat-soft}

For semi-inclusive quantities like the non-global observables
of~\cite{Banfi:2002hw}, in which one sums over a given, limited phase
space volume of soft gluons, it is natural to ask for the dependence
on that phase space boundary. Unlike a direct calculation of the
average in Eq.~\eqref{eq:prob-from-func-all} this should lead to a
tractable result that exhibits new structure. In fact, I shall
demonstrate that one recovers the generalization to the BMS equation
suggested above.

To arrive at an evolution equation, I will simply take a derivative
w.r.t. to the phase space boundary of the emitted gluons in the
semi-inclusive probability~\eqref{eq:prob-from-func-all}. This is in
direct correspondence with the procedure used in BMS. Here it
is essential that one includes the phase space boundary into the
definition of $M$. Then the result is most transparently displayed
using the following simple relationship, based on functional
differentiation and Legendre transformation in the case of a Gaussian
action (``free theory''):
\begin{align}
  \label{eq:average-Legendre}
   \big\langle W[\Xi] \big\rangle_{\Xi}
   = & N \int D[\Xi]\ W[\Xi]\
   e^{-\frac{\Xi^t M^{-1} \Xi}{2}}
   = 
    W[\frac{\delta}{i\, \delta  {\cal J}}] 
    e^{-\frac{{\cal J}^t M {\cal J}}{2}}\Big\vert_{{\cal J}=0}
    = e^{-\frac{\frac{\delta}{i \delta \Xi_0} M 
        \frac{\delta}{i \delta \Xi_0}}{2}} W[\Xi_0]\Big\vert_{\Xi_0=0}
\end{align}
for any functional $W[\Xi]$. Here notation has been condensed even
further, with all momenta, integration signs and measures
suppressed.\footnote{For compactness I will do so throughout this
  section. Where indices and momenta are shown it is with the
  understanding that there will be integration conventions for
  repeated momenta with $d\Phi_k$ as the measure.}  Legendre machinery
is used for the last equality sign. The ``classical field,'' $\Xi_0 =
i M {\cal J}$, vanishes at ${\cal J}=0$.

The canonical example for $W[\Xi]$ here of course is
\begin{equation}
  \label{eq:W-example}
  \big\langle W[\Xi] \big\rangle_{\Xi}\to  
  \big\langle{\sf A}_{p\,q}^{ij}[U,\Xi] 
  {\sf A}_{p\,q}^{i j}[U,\Xi]^\dagger\big\rangle_{\Xi}
\ ,
\end{equation}
which, by construction, is the finite $N_c$ generalization of
$G^{(\text{real})}_{a b}$.

For this example one immediately realizes that the exponential of the
second order differential operator on the right hand side will perform
precisely the ``sewing'' implemented by the Gaussian weight in
Eq.~\eqref{eq:prob-from-func-all} as long as $M$ is diagonal in
energies as in \eqref{eq:M-BMS}. In this case, the energy ordering
ensures that
\begin{equation}
  \label{eq:sewing-step}
  \frac{1}{2}\frac{\delta}{i \delta \Xi_0} M 
        \frac{\delta}{i \delta \Xi_0}{\sf A}_{p\,q}^{i j}[U,\Xi_0]
 {\sf A}_{p\,q}^{i
  j}[U,\Xi]^\dagger = 
\Big(\frac{\delta}{i \delta \Xi_0}{\sf A}_{p\,q}^{i j}[U,\Xi_0] \Big)
 M 
        \Big(\frac{\delta}{i \delta \Xi_0}
{\sf A}_{p\,q}^{i
  j}[U,\Xi]^\dagger \Big)\ .
\end{equation}

Let me now carry out the derivative with respect to the phase space
boundary in analogy with the derivation of the BMS equation sketched
in Sec.~\ref{sec:analogy}. With $M=M(E)$ a function of the phase space
boundary, one immediately finds a general expression for the
logarithmic $E$-derivative of the above expectation value that reads
\begin{equation}
  \label{eq:Legendre-E-der}
 E\partial_E \ \big\langle W[\Xi] \big\rangle_{\Xi}(E) 
= - \frac{1}{2}e^{-\frac{\frac{\delta}{i \delta \Xi_0} M(E) 
        \frac{\delta}{i \delta \Xi_0}}{2}} \
    \frac{\delta}{i \delta \Xi_0} E\partial_E M(E) 
        \frac{\delta}{i \delta \Xi_0}\ 
        W[\Xi_0]\Big\vert_{\Xi_0=0}
\ .
\end{equation}

The main point about this seemingly trivial exercise is its use in
conjunction with what is already known about functional derivatives of
the amplitudes of interest:
Eqns.~\eqref{hard-variation},~\eqref{eq:qqb-qqbg-rel}. Note that
$\partial_E M(E)$ will force the variations in the factor taken down
from the exponential to be at the phase space boundary. Choosing
$W[\Xi]= {\sf A}_{p\,q}^{i j}[U,\Xi] {\sf A}_{p\,q}^{i
  j}[U,\Xi]^\dagger$, the square of the generating functional of
$q\Bar q\, g^n_{\text{soft}}$ amplitudes, this will lead to the
appearance of the square of the generating functional of $q\Bar q g\,
g^n_{\text{soft}}$ on the right hand side as follows
from~\eqref{eq:qqb-qqbg-rel},~\eqref{eq:Xivar}.

This marks an important difference compared to the BMS case at finite
$N_c$: the equation does not close. Instead higher and higher hard
correlators will enter: Clearly, an evolution equation for the $q\Bar
q g\, g^n_{\text{soft}}$ amplitudes will then couple in turn to the
corresponding object for $q\Bar q g^2\, g^n_{\text{soft}}$ amplitudes
and so forth.  One is faced with an infinite hierarchy. All quantities
encountered in this hierarchy can be defined in complete analogy to
the examples presented in~\eqref{eq:exp-amp} and \eqref{eq:qqbg}. One
simply uses the desired combination of $U$-factors as the ``bare''
term in the tower of amplitudes one is interested in.

This implies that firstly, it is not sufficient to consider only the
evolution equation for $q\Bar q g^n_{\text{soft}}$ amplitudes alone.
Instead one has to capture the complete infinite coupled hierarchy of
equations in one. Secondly, it indicates that such a step is possible,
since the objects of interest have already been identified and indeed
can be written down in a general form. The evolution equations then
follow from~\eqref{eq:Legendre-E-der}.

I therefore proceed to define the (tree level) generating functional
based on $m$ hard particles, a general ``antenna pattern,'' by
\begin{equation}
  \label{eq:m-hard-particles}
  {\sf A}_{p_1 \ldots p_m}^{i_1
  \ldots i_n}[U,\Xi] {\sf A}_{p_1\ldots p_m}^{i_1 \ldots i_n}[V,\Xi]^\dagger
  = {\sf U}[\Xi,U]
  \
  (UV^\dagger)_{p_1}^{(\dagger)}\otimes  \ldots  
  \otimes (UV^\dagger)_{p_m}^{(\dagger)}
  \ 
  {\sf U}[\Xi,V]^\dagger \Big\vert_{U=V}
\ .
\end{equation}
The factor $ (UV^\dagger)_{p_1}^{(\dagger)}\otimes \ldots \otimes
(UV^\dagger)_{p_m}^{(\dagger)}$ in this expression generalizes the
expression for the bare particles of the $q\Bar q$ case (the ${\sf
  A}_{p\,q}^{i j}[U,\Xi] {\sf A}_{p\,q}^{i j}[U,\Xi]^\dagger$ from
above) from $2$ to $m$ hard legs, and the ${\sf U}[\Xi,U] \ \ldots \ 
{\sf U}[\Xi,V]^\dagger \Big\vert_{U=V}$ implement the showering.
I have given only $q$ and $\Bar q$ factors as any
gluon can be written in terms of these as $\Tilde{(UV^\dagger)}_k^{a
  b} = 2\tr( t^a (UV^\dagger)_k^\dagger t^b (UV^\dagger)_k)$.

One can now study the evolution of the expectation value of a given
object of the type~\eqref{eq:m-hard-particles} or directly study a
generating functional for {\em all} such objects in one go. The latter
is given by
\begin{align}
  \label{eq:genfunc}
  {\sf W}[j^\dagger,j] := & \Big\langle
  {\sf U}[\Xi,U] \
  e^{i 2\tr j^\dagger U V^\dagger +
  i 2\tr (U V^\dagger)^\dagger j}
  \
  {\sf U}[\Xi,V]^\dagger   
  \Big\rangle_{\Xi}\Big\vert_{U=V}
\end{align}
with integrals in the source exponents understood.  Appropriate
variations with respect to $j$ and $j^\dagger$ at $j=0$ will then
select a given tower of $\text{in}\to\text{hard}\, g^n_{\text{soft}}$
probabilities entering the hierarchy. Since, besides the phase space
constraints contained in $M$, it is the particle content of the bare
terms that define the amplitudes, a notation that emphasizes this is
needed. To this end I will write
\begin{align}
  \label{eq:UV-average}
  \frac{\delta}{i\delta j^{(\dagger)}_{p_1}} \ldots  
  \frac{\delta}{i\delta j^{(\dagger)}_{p_m}}  
  {\sf W}[j^\dagger,j]
  = &\Big\langle
  {\sf U}[\Xi,U] \
  (UV^\dagger)_{p_1}^{(\dagger)}\otimes  \ldots  
  \otimes (UV^\dagger)_{p_m}^{(\dagger)}
  \
  {\sf U}[\Xi,V]^\dagger   
  \Big\rangle_{\Xi}\Big\vert_{U=V}
\nonumber \\ 
  =: &
  \Big\langle \ (UV^\dagger)_{p_1}^{(\dagger)}\otimes  \ldots  
  \otimes (UV^\dagger)_{p_m}^{(\dagger)}\
  \Big\rangle_{UV^\dagger}^{\text{real}}
\end{align}
where the $UV^\dagger$ average in the second line is to be taken with
a yet unknown weight that reproduces the l.h.s.. The idea is to
interpret the soft gluon average as an average over $UV^\dagger$
configurations. This is completely legal, if somewhat formal as long
as the only definition of the weight of the average is
through~\eqref{eq:UV-average}.

Nevertheless, it is this reinterpretation that will allow to make
contact with the Fokker-Planck formulation of
Sec.~\ref{sec:from-jimwlk-bk} and that is all that is really needed.

For $m=2$ and an adapted choice for the color structure one encounters
\begin{equation}
  \label{eq:G-at-m=2}
  \langle 
  \tr((UV^\dagger)^\dagger_p (UV^\dagger)_q)/N_c 
  \rangle_{UV^\dagger}^{\text{real}} 
  =: G_{a b}^{(\text{real})}(E)
\end{equation}
and thus provides the precise definition of what was still a bit vague
in~\eqref{eq:interpretation}.  Eq.~\eqref{eq:UV-average} is of course
much more general than that: it defines the real emission part for a
general antenna pattern, as these will all be needed for the evolution
equations.  Inclusion of virtual contributions will be discussed
below.  Note that the eikonal lines appearing in~\eqref{eq:UV-average}
have the interpretation of a product $UV^\dagger$ of contributions
from both the amplitude and the conjugate amplitude.

With~\eqref{eq:average-Legendre} one then has
\begin{align}
  \label{eq:funcform}
   {\sf W}[j^\dagger,j] =  e^{-\frac{\frac{\delta}{i \delta \Xi_0} M(E) 
        \frac{\delta}{i \delta \Xi_0}}{2}}
{\sf U}[\Xi_0,U]\
  e^{i 2\tr j^\dagger U V^\dagger +
  i 2\tr (U V^\dagger)^\dagger j}
\ 
{\sf U}[\Xi_0,V]^\dagger
\Big\vert_{U=V}
=   \big\langle   e^{i 2\tr j^\dagger U V^\dagger +
  i 2\tr (U V^\dagger)^\dagger j} \big\rangle_{UV^\dagger}^{\text{real}}
\ .
\end{align}
This demonstrates that $ {\sf W}[j^\dagger,j] $ is in complete analogy
with (the real emission part of) $\Bar Z[J,J^\dagger]$ in the
derivation of the JIMWLK equation~\cite{Weigert:2000gi}.

To find the evolution equation for this (meta-) functional and with it
the infinite hierarchy of evolution equations alluded to above, all
that is left to do, is to put together
Eqns.~\eqref{eq:Legendre-E-der},~\eqref{eq:sewing-step},
and~\eqref{hard-variation} to get the dependence on phase space
boundaries for arbitrary $UV$-correlators (bare $m$-jet probabilities).
This yields the r.h.s. for the real emission contribution:
\begin{align}
  \label{eq:real-emission}
    & e^{-\frac{\frac{\delta}{i \delta \Xi_0} M(E) 
        \frac{\delta}{i \delta \Xi_0}}{2}}
{\sf U}[\Xi_0,U]  
\nonumber \\ & 
\Big\{
  g\, J_{l k}^{\mu} \Tilde U^{a b}_{k} 
  i{\Bar\nabla}^a_{U_l} \Big\} [E\partial_E M(E)]_{l l'\,\mu\mu'}^{b b'}
  \Big\{
  g\, J_{l' k'}^{\mu'} \Tilde V^{a' b'}_{k'} 
  i{\Bar\nabla}^a_{V_l'} \Big\} \
  e^{i 2\tr j^\dagger U V^\dagger +
  i 2\tr (U V^\dagger)^\dagger j}
\nonumber \\ & \ {\sf U}[\Xi_0,V]^\dagger \Big\vert_{U=V}
\ .
\end{align}
To arrive at a meaningful answer one still has to supplement virtual
corrections.  They are correctly incorporated as usual, by simply
subtracting the corresponding term proportional to the original
transition probability. This was done in the derivation of the BMS
equation and it also applies to the full hierarchy. For concreteness,
and without loss of generality, I will demonstrate how to achieve this
for the case of $m$ hard particles shown in Eq.~\eqref{eq:UV-average}.


To understand the constraints on the virtual corrections, let me first
study the real emission contribution in some more detail.  Omitting,
for the moment, the operator that implements the soft gluon shower and
the expectation value in the expression for the real emission part,
\[e^{ -\frac{\frac{\delta}{i \delta \Xi_0} M(E)
    \frac{\delta}{i \delta \Xi_0}}{2}} {\sf U}[\Xi_0,U] \ldots {\sf
  U}[\Xi_0,V]^\dagger\Big\vert_{U=V}\ ,
\]
one has to consider
\begin{align}
  \label{eq:real-emission-1} 
\Big\{
  g\,J_{l k}^{\mu} \Tilde U^{a b}_{k} 
  i{\Bar\nabla}^a_{U_l} \Big\}
  [ E\partial_E M(E)]_{l l'}\delta^{b b'}g_{\mu\mu'}
  \Big\{ 
  g\, J_{l' k'}^{\mu'} \Tilde V^{a' b'}_{k'} 
  i{\Bar\nabla}^a_{V_l'} \Big\} \
  (UV^\dagger)_{p_1}^{(\dagger)}\otimes  \ldots  
  \otimes (UV^\dagger)_{p_m}^{(\dagger)}
\end{align}
and evaluate the variations. Inserting the BMS choice~\eqref{eq:M-BMS}
for $M$, it is easy to collect all the factors and to make the $k$
integration explicit (I keep the integration convention for $l$ and
$l'$ for readability):
\begin{align}
  \label{eq:get-kernel}
  \eqref{eq:real-emission-1} = &
  \int \frac{d\omega_k}{\omega_k} E \delta(E-\omega_k) \frac{d\Omega_k}{4\pi}
  \frac{\alpha_s}{\pi} w_{l l'}(k)   (\Tilde{UV}^\dagger)^{a b}_k
   i{\Bar\nabla}^a_{U_l} i{\Bar\nabla}^b_{V_l'} \
  (UV^\dagger)_{p_1}^{(\dagger)}\otimes  \ldots  
  \otimes (UV^\dagger)_{p_m}^{(\dagger)}
\nonumber \\ = &
\int\!\frac{d\Omega_k}{4\pi}
  \frac{\alpha_s}{\pi} w_{l l'}(k)  (\Tilde{UV}^\dagger)^{a b}_k  
   i{\Bar\nabla}^a_{U_l} i{\Bar\nabla}^b_{V_l'} \
  (UV^\dagger)_{p_1}^{(\dagger)}\otimes  \ldots  
  \otimes (UV^\dagger)_{p_m}^{(\dagger)}\Big\vert_{\omega_k=E}
  \ .
\end{align}
In this expression $\omega_l$ and $\omega_l'$ will be made hard when
the functional derivatives hit the $U$ and $V$. The energy part of the
corresponding phase space integrals will be eliminated at the same
time. In summary one will end up with 3 ``hard'' momenta $k,l,l'$.

Carrying out the differentiations a sum over pairings will emerge,
with different color structures depending if one hits $q$ or $\Bar q$
lines. Concentrating on the individual terms in this sum, it is
obvious that one has to deal with 4 different combinations,
corresponding to the pairings $qq$, $q\Bar q$, $\Bar q q$, and $\Bar
q\Bar q$. This is a structure well known from the the derivation of
both the Balitsky hierarchy~\cite{Balitsky:1996ub} and the JIMWLK
equation in~\cite{Weigert:2000gi}.  There I had first to unearth the
invariant vector fields from a diagrammatic calculation that has
started off from an entirely different setup. Since this tool is
already available, I will not go through the line of argument used
there, but instead make use of the unifying properties of these
operators right away without exploring these structures explicitly.

This may be done by noting that
Eq.~\eqref{eq:real-emission-1} allows for a very elegant rewrite:
\begin{align}
  \label{eq:real-rewrite}
  \text{~\eqref{eq:real-emission}} = &
 \frac{\alpha_s}{2\pi} w_{p q}(k)u(k) (\Tilde{UV^\dagger})^{a b}_k 
\big( i \Bar\nabla_{(UV^\dagger)_p}^a  i\nabla^b_{(UV^\dagger)_q} 
+i \Bar\nabla_{(UV^\dagger)_q}^a  i\nabla^b_{(UV^\dagger)_p}\big)\ 
  (UV^\dagger)_{p_1}^{(\dagger)}\otimes  \ldots  
  \otimes (UV^\dagger)_{p_m}^{(\dagger)}
\ .
\end{align}
Eq.~\eqref{eq:real-rewrite} follows from a direct correspondence of
$i\Bar\nabla^a_{(UV^\dagger)_k}$ with $i\Bar \nabla^a_{U_k}$ acting on
the amplitude and $i\nabla^a_{(UV^\dagger)_k}$ with $i\Bar
\nabla^a_{V_k}$ acting on the complex conjugate factor.  In
Eq.~\eqref{eq:real-rewrite} it is possible to anticipate that the
energies will all be made hard by functional differentiation on the
hard factors $(UV^\dagger)_{p_i}$ and thus to go back to the original
definition of the variations without the energy $\delta$-function.
This allows to use an integration convention that employs the solid
angle integrations only and a symmetric treatment of $k,p,q$ in this
equation. Reinstating the shower-operators and the expectation value,
one arrives at the real emission part of the evolution equation
\begin{align}
  \label{eq:real-evolution}
 \Big\langle 
 \frac{\alpha_s}{2\pi} w_{p q}(k)u(k) (\Tilde{UV^\dagger})^{a b}_k 
\big( i \Bar\nabla_{(UV^\dagger)_p}^a  i\nabla^b_{(UV^\dagger)_q} 
+i \Bar\nabla_{(UV^\dagger)_q}^a  
i\nabla^b_{(UV^\dagger)_p}\big)\ 
   (UV^\dagger)_{p_1}^{(\dagger)}\otimes  \ldots  
  \otimes (UV^\dagger)_{p_m}^{(\dagger)}
  \Big\rangle_{UV^\dagger}(E) 
\ .
\end{align}

To discuss virtual corrections one needs to consider the physical
process at hand. While the real emissions are confined to the {\em
  inside} regions and hence carry a factor $u(k)$, virtual corrections
appear everywhere and therefore have no such factor.  Moreover the
factor $(\Tilde UV^\dagger)_k^{a b}$ will be replaced by $\delta^{a
  b}$ and and the vector fields will have to act twice {\em within}
either the amplitude or its complex conjugate, leading to
$i\nabla_{(UV^\dagger)_p}^a i\nabla^a_{(UV^\dagger)_q}$ and an
analogous barred contribution instead of the mixed ones in
Eq.~\eqref{eq:real-evolution}. For symmetry reasons one therefore
expects the virtual corrections to read
\begin{align}
  \label{eq:virtual-rewrite}
 \Big\langle \frac{\alpha_s}{2\pi} w_{p q}(k) 
 \big( i \nabla_{(UV^\dagger)_p}^a  i\nabla^a_{(UV^\dagger)_q} 
+i \Bar\nabla_{(UV^\dagger)_q}^a  i\Bar\nabla^a_{(UV^\dagger)_p}\big)\ 
  (UV^\dagger)_{p_1}^{(\dagger)}\otimes  \ldots  
  \otimes (UV^\dagger)_{p_m}^{(\dagger)}
  \Big\rangle_{UV^\dagger}(E) 
\end{align}
where the overall sign and normalization is fixed by real virtual
cancellation: only with this choice do they add up to an infrared
finite Fokker-Planck Hamiltonian, which happens to coincide with the
expression conjectured in Sec.~\ref{sec:analogy}. Indeed, the operator
appearing in the sum of~\eqref{eq:real-evolution}
and~\eqref{eq:virtual-rewrite} is nothing but $-H_{\text{ng}}$ of
Eq.~\eqref{eq:H_ng-0} with the $f^{(i)}$ completely determined. In
particular, there is no room for any $N_c$ dependence of the
coefficients from this argument.

To summarize, it has been shown that
\begin{equation}
  \label{eq:gen-evol}
  E \partial_E\ \Big\langle   
  e^{i 2\tr j^\dagger U V^\dagger +
  i 2\tr (U V^\dagger)^\dagger j} \Big\rangle_{UV^\dagger}(E)
  = 
- \Big\langle \  H_{\text{ng}} \ 
  e^{i 2\tr j^\dagger U V^\dagger +
  i 2\tr (U V^\dagger)^\dagger j}\ \Big\rangle_{UV^\dagger}(E)
\ .
\end{equation}
The only argument still missing is about how to reverse the step
leading from~\eqref{eq:JIMWLK} to~\eqref{eq:OU-evo}.

This follows from the fact that the identity holds for arbitrary $j$
and $j^\dagger$, i.e.  the fact that the above is completely
independent of the type of correlator considered. The result is, as
advertised in Sec.~\ref{sec:from-jimwlk-bk}, the evolution equation
\begin{equation}
  \label{eq:new-FP}
  E\partial_E\ \Hat Z_E[UV^\dagger] = -H_{\text{ng}}\ \Hat Z_E[UV^\dagger]
\end{equation}
for the weight $Z_E[UV^\dagger]$ used to define the averages $\langle
\ldots \rangle_{UV^\dagger}(E)$. This completes the derivation of
Eq.~\eqref{eq:FP-ng}.


I will close this section with the explicit expressions for the
non-factorized version of the BMS equation as it emerges from this
discussion, to point out how the $N_c$ limit comes about and to
illustrate once more the real-virtual cancellations with a concrete
example. As this is concerned with emissions from a bare $q\Bar q$
jet, one needs the operator and probabilities
\begin{subequations}
\begin{align}
  \label{eq:S-2-def} 
\Hat G_{p q} := & \tr((UV^\dagger)^\dagger_p (UV^\dagger)_q)/N_c
\\
G_{p q}(E) := &  \langle \Hat G_{p q} \rangle_{UV^\dagger}(E)  =
\frac{1}{N_c}
\frac{\delta}{i \delta (j_p)_{i j} } \frac{\delta}{i \delta (j_p)_{i j} } 
{\sf W}_E[j^\dagger,j]
\ .
\end{align}  
\end{subequations}
The resulting equation then is
\begin{align}
  \label{eq:two-point-to3point-minus-virt-1}
  E\partial_E   & \langle  
  \tr((UV^\dagger)^\dagger_p (UV^\dagger)_q)/N_c\rangle_{UV^\dagger}(E)
\nonumber \\ = & 
 \int\!\frac{d\Omega_k}{4\pi} \frac{\alpha_s}{\pi} w_{p q}(k) \ 
\Big\langle 
u(k)\Tilde {(UV^\dagger)}_k^{a b}\ 2 
\frac{\tr(t^a (UV^\dagger)_p t^b (UV^\dagger)_q)}{N_c} 
- 2 C_{\text{f}} 
\frac{\tr((UV^\dagger)^\dagger_p (UV^\dagger)_q)}{N_c}  
\Big\rangle_{UV^\dagger}(E) 
\ .
\end{align}
The first term on the r.h.s., carrying the factor $u(k)$, originates
from the real contribution, the second from the virtual one.
By~\eqref{eq:Fierz} this is equivalent to
\begin{subequations}
  \label{eq:general-minus-virt}
\begin{align}
   E\partial_E  G_{p q}(E) 
= &   \int\!\frac{d\Omega_k}{4\pi}\frac{\alpha_s}{\pi} w_{p q}(k)
   \Big\langle  u(k)\big(\Hat G_{p k} \Hat G_{k q} N_c 
   - \frac{\Hat G_{p q}}{N_c}\big) 
   -2 C_{\text{f}} \Hat G_{p q} \Big\rangle (E)
 \\
= & \int\!\frac{d\Omega_k}{4\pi}\Bar\alpha_s w_{p q}(k) 
\Big\langle u(k)\big(\Hat G_{p k} \Hat G_{k q} - \Hat G_{p q} \big)
 -\frac{2 C_{\text{f}}}{N_c}(1-u(k))\Hat G_{p q} \Big\rangle (E)
\ ,
\end{align}  
\end{subequations}
which in the large $N_c$ limit factorizes into the analogues of the BK
that triggered these explorations, the BMS equation
Eqns.~\eqref{eq:BMS-1} and~\eqref{eq:BMS-2} respectively. To see this
explicitly, use~\eqref{eq:average-Legendre} to represent the averaging
involved in
\begin{equation}
  \label{eq:G-soft-aver}
  G_{p q}(E) =  \langle \Hat G_{p q} \rangle_{UV^\dagger}(E)  
  =\langle  {\sf A}_{p\,q}^{i j}[U,\Xi] 
  {\sf A}_{p\,q}^{i j}[U,\Xi]^\dagger \rangle_\Xi
\end{equation}
and then repeat the argument of\cite{Fiorani:1988by}, Eq.~(50). It is
worth emphasizing that the real-virtual cancellation argument of BMS
restated in Sec.~\ref{sec:analogy} applies already to
Eq.~\eqref{eq:general-minus-virt} itself, without any reference to the
factorization argument. Infrared finiteness in the Sudakov term
follows directly from the argument of BMS while in the other term it
is the definition of $\Hat G$ that saves the day. This provides an
illustration of how the more general reasoning above will lead to
infrared finite results via real-virtual cancellation in {\em any}
evolution equation for color singlet objects contained in the new
evolution equation.

\section{Conclusions}
\label{sec:conclusions}

The main result of this paper is the evolution
equation~\eqref{eq:FP-ng} that generalizes the BMS equation to finite
$N_c$. The result is functional and represents an infinite hierarchy
of coupled equations just as its small $x$ analogue, the JIMWLK
equation. As in this case, a Langevin formulation has been derived
that hopefully can be implemented efficiently as a numerical
simulation.

More subtle benefits lie in the method used to prove the result.  This
was based on a formulation of strongly ordered soft emission
amplitudes using shower operators that can be used to generate soft
gluon clouds around any hard seed. The ordered nature of these is the
main calculational benefit that eventually allowed the derivation of
the new evolution equation. This concept should prove useful in most
of the new applications briefly touched upon below.

The amplitudes were then translated into probabilities and several
forms for taking the expectation values were given. Each of these has
its own advantage as shown by their use in the derivation of the
evolution equation. This in the end has allowed to derive this equation
by simply differentiating the expressions found with surprisingly
simple manipulations. The simplicity of these steps allows the
modification of the equation to applications with other global and
non-global jet observables.

By the close analogy with JIMWLK and BK, some of the results obtained
in the asymptotic regime, such as the presence of the saturation
scale clearly seem to map on universal features of the jet evolution,
such as the existence of $\theta_{\text{crit}}$ discussed by BMS.
Further explorations along these lines appear to be promising.

Conversely, it should be possible to use the strategy employed here to
rederive the JIMWLK equation. In this case, quantities that are less
inclusive than the dipole cross section in DIS should become
accessible. Diffraction or the formulation of observables that focus
in on small impact parameters would emerge by formulating appropriate
phase space constraints. The definition of observables that are
infrared safe regardless of impact parameter and gluon density
encountered is another item on this list. Such a step will certainly
come at the price of an additional term in the evolution equation just
like the Sudakov term in BMS, but would put applications of that
equation on solid ground even for small targets like the proton. This
should allow a new level of insight for the application of saturation
ideas to HERA data and ultimately also help with applications to RHIC
and LHC physics.


\paragraph{Acknowledgements:} 
I am grateful to Andreas Sch\"afer and Vladimir Braun for a thorough
discussion of content and presentation and want to thank Einan Gardi,
Andreas Freund, Mrinal Dasgupta and Kari Rummukainen for a string of
communications and discussions on various subtopics.  I also want to
acknowledge partial support by  BMBF.

\appendix

\section{Color algebra of the induction step}
\label{sec:color-algebra-induct}

The color algebra pattern can be best understood by looking at the
small values for $n$. This uncovers the pattern for the general case.
Notation below is simplified in order to expose the structure of terms:
  \begin{itemize}
  \item n=0: starting from $U^\dagger_p U_q$ one arrives at
    \[
    -J_{p a} U^\dagger_p t^a U_q + J_{a q} U^\dagger_p t^a U_q 
    = U^\dagger_p t^a U_q A_{p q}(k)
    \]
  \item n=1: starting from $(U^\dagger_p t^{a_1} U_q) \Tilde U^{a_1
      b_1}$ one finds
  \begin{align*}
  -J_{p a} U^\dagger_p t^a t^{a_1} U_q \Tilde U^{a_1 b_1}
  + J_{a_1 a} (U^\dagger_p [t^a,t^{a_1}] U_q) \Tilde U^{a_1 b_1}
  + J_{q a} U^\dagger_p t^{a_1} t^a U_q) \Tilde U^{a_1 b_1}
\\
  =  ( J_{a_1 a}-J_{p a})U^\dagger_p t^a t^{a_1} U_q  U^{a_1 b_1}
  + (J_{q a}-J_{a_1 a})U^\dagger_p t^{a_1} t^a U_q \Tilde U^{a_1 b_1}
\end{align*}
\item n=2: starting from $(U^\dagger_p t^{a_1} t^{a_2} U_q) \Tilde
  U^{a_1 b_1} \Tilde U^{a_2 b_2}$ one obtains
  \begin{align*} 
    & -J_{p a} (U^\dagger_p t^a t^{a_1} t^{a_2} U_q \Tilde U^{a_1 b_1}
    \\ &
   + J_{a_1 a} (U^\dagger_p [t^a, t^{a_1}] t^{a_2} U_q) \Tilde U^{a_1 b_1} 
   + J_{a_2 a}(U^\dagger_p t^{a_1} [t^a,t^{a_2}] U_q) \Tilde U^{a_1 b_1}    
   \\ &
   +  J_{q a} (U^\dagger_p t^{a_1} t^{a_2} t^a U_q)) \Tilde U^{a_1 b_1}
  \end{align*}
\end{itemize}
The general step is now obvious: to collect terms with the same color
structure, the first term (from the $U^\dagger$) pairs with the first
term of the first commutator, then iteratively the second term of one
commutator with the first term of the next. This continues all the way
through until the last remaining term of the last commutator pairs up
with the term of the $U$ factor. In all the pairings, the signs will
alternate, yielding the $A_{l l'}(k)$.  Setting the labels accordingly
then leads to the induction step given in
Eq.~\eqref{eq:fioriani-n-step}.


\providecommand{\href}[2]{#2}\begingroup\raggedright\endgroup

\end{document}